\documentclass[preprint,12pt]{elsarticle}
\usepackage{graphicx}
\usepackage{pgfplotstable}
\usepackage{floatrow}
\usepackage{longtable}
\usepackage{amsmath}
\usepackage{amssymb}
\usepackage{amsmath}
\usepackage{amsthm}
\usepackage{subfig}
\usepackage{hyperref}
\hypersetup{
    colorlinks=true,
    linkcolor=blue,
    filecolor=magenta,      
    urlcolor=cyan,
}
\usepackage{lineno}
\usepackage{subfig}
\newcommand{\AAA}{\mathcal{A}}
\newcommand{\SSS}{\mathcal{S}}
\newcommand{\OOO}{\mathcal{O}}
\newcommand{\DDD}{\mathcal{D}}
\newcommand{\LLL}{\mathcal{L}}
\newcommand{\NNN}{\mathcal{N}}
\newcommand{\TTT}{\mathcal{T}}
\newcommand{\WWW}{\mathcal{W}}
\newcommand{\PPP}{\mathcal{P}}
\usepackage[misc]{ifsym}

\journal{Transportation Research Part C}

\begin{document}

\begin{frontmatter}


\title{Optimizing timetable and network reopen plans for public transportation networks during a COVID19-like pandemic}

\author[berkeley]{Yiduo Huang}
\author[berkeley]{Zuojun Shen\corref{fulladdress}}
\ead{maxshen@berkeley.edu}
\cortext[fulladdress]{416 McLaughlin Hall, Berkeley, CA 94720, United States}
\address[berkeley]{Berkeley, California, United States}

\begin{abstract}
The recovery of the public transportation system is critical for both social re-engagement and economic rebooting after the shutdown during pandemic like COVID-19. In this study, we focus on the integrated optimization of service line reopening plan and timetable design. We model the transit system as a space-time network. In this network, the number of passengers on each vehicle at the same time can be represented by arc flow. We then apply a simplified spatial compartmental model of epidemic (SCME) to each vehicle and platform to model the spread of pandemic in the system as our objective, and calculate the optimal open plan and timetable. We demonstrate that this optimization problem can be decomposed into a simple integer programming and a linear multi-commodity network flow problem using Lagrangian relaxation techniques. Finally, we test the proposed model using real-world data from the Bay Area Rapid Transit (BART) and give some useful suggestions to system managers.
\end{abstract}


\end{frontmatter}
%\linenumbers
\section{Introduction}
\label{S:1}
In response to the COVID-19 outbreak, the Center of Disease Control and Prevention (CDC) has promulgated guidance, including the practice of social distancing. The state of California also made intensive efforts and became the first state to issue a mandatory shelter-in-place order, which has largely mitigated the spread of the disease. Consequently, both the demand and supply for public transport decreased drastically from March to September.

As the situation evolves, regular activities will resume in the foreseeable future. Public transit, which is an integral part of personal lives, will become the first to face the sheer volume of crowds. Great concerns along with transmission relapse may be raised again when quantifying a crowded transportation system's specific risk. It is worth noting that although many efforts have been made to model the spread of the disease and the public transportation network design problem has been thoroughly studied in the transportation literature, no clear guidance is available to public transit agencies during the recovery phase. In contrast, the infection risk of a pandemic like COVID-19 is not considered in the existing network planning and scheduling methods. Furthermore, safety measures such as social distancing and vehicle disinfection change the public transportation system's working procedure. These factors need to be considered when drafting the recovery plan.

Therefore, a reopening design method considering the risk of new infection is needed for public transportation and becomes more urgent in the coming months. Taking the active social distancing order and the risk of new infection into consideration, we address the following problems simultaneously: (1) Network reopening design: how to re-gauge the network and determine which lines and stations to open in the given region. We need to decide if we can open each line or station at the beginning of the day; (2) Timetable design: how to design a timetable to satisfy all demands and control the risk of new cases. We need to decide when we should dispatch one train on line $l$, assuming a fixed travel time. We say one train on line $l$ is dispatched at time $t$ if one train of line $l$ leaves its first stop and start a new run. We call one train operation a `dispatch' and we use term `run' and `dispatch'  interchangeably in this paper.

\subsection{Pandemic and travel}
\label{S:1.1}
The COVID-19 pandemic has dramatically impacted the global economy in 2020, with 167K deaths and $>5$M confirmed cases in the US alone by September 2020 \cite{JHU}.

The transmission mechanism of COVID-19 and other similar epidemics have been studied by public health researchers from January to August, and new results are being consistently obtained. It is already known that COVID-19 can be transmitted directly from person to person through respiratory droplets \cite{CDC2020a}, and it is likely that “a person can get COVID-19 by touching a surface or object that has the virus on it” \cite{CDC2020b}. The disease is highly contagious, with a reproduction rate of $R_0$ estimated to be $2.7$ to $5.7$ in Wuhan, China \cite{Sanche2020} in September 2019 and 6.4 in New York State \cite{Ives2020} in March 2020.

To model the transmission of COVID-19 or other pandemic in a public transit network, the spatial nature of the pandemic's spread should be captured explicitly. There are plenty of studies on modeling the spread of pandemics in the mathematical epidemiology literature (see books such as \cite{Allen2008}, \cite{Brauer2019}, and \cite{Martcheva2015}). Generally, the pandemic spread can be modeled using a network where each node represents one location (a city, province, or block). Inside each node, the pandemic spread according to differential or difference equations, similar to the Susceptible, Infectious, or Recovered (SIR) model. 

During the COVID-19 pandemic, researchers have further developed these models to evaluate COVID-19 travel policies. Chinazzi \cite{Chinazzi2020} studied the impact of travel restrictions in China. They used the global epidemic and mobility model (GLEAM) and model regions as nodes in a metapopulation network based on air transportation data.
Jia et al.\cite{Jia2020} developed a spatiotemporal `risk source' model to assess the risk of COVID-19 community transmission using the actual movement data of cell phone users. They modeled the infection based on a multiplicative exponential model and predicted the number of new infections in cities or provinces outside Wuhan.
Li et al.\cite{Li2020} simulated the spatiotemporal dynamics of infection and inferred important parameters using an iterative filtering-ensemble adjustment Kalman filter framework. They concluded that the transmission rate $\beta$ in spatial compartmental models is $1.12$ ($95\%$ CI $[1.06,1.19]$). This parameter captures the rate at which an infected agent infects others and plays an important role in our model.

Based on these models, operations management and economics researchers have studied the optimal lockdown or reopen decision making aspects during the pandemic. Birge et al.\cite{Birge2020} proposed a spatial compartmental model of epidemics (SCME) that was an extension of the works of \cite{Allen2008}\cite{Brauer2019}\cite{Martcheva2015}.They optimized the neighborhood travel ban/reopen plan policy in New York City. 
\cite{Kaplan2020} proposed a decision-making model implemented at Yale University. Alvarez et al.\cite{Alvarez2020} attempted to find the optimal lockdown policy to minimize the loss. Other researchers such as \cite{Acemoglu2020}\cite{Gershon2020}\cite{Glover2020} used heterogeneous Susceptible-Infectious-Removed (SIR) or Susceptible-Exposed-Infectious-Removed (SEIR) models to make critical decisions on the lockdown period or medical supply distribution during the pandemic.

\subsection{Pandemic and public transit}
\label{S:1.2}

The COVID-19 pandemic has heavily impacted public transit systems in the US. For example, in the bay area, Bay Area Rapid Transit (BART) implements social distancing procedures, and masks are mandatory. They use hospital-grade disinfectants to clean the carts multiple times a day \cite{BART2020b}. AC transit, who provide bus services to the east San Francisco Bay Area, have modified the timetable and canceled some line services \cite{AC2020}. Ridership in New York dropped by 92\% in mid-April, and night services are no longer available \cite{Frost2020}. Similar measures have been implemented by different transit agencies across the US \cite{APTA2020}\cite{CDC2020c}.

To date (May 2021), there is no confirmed evidence of COVID-19 transmission via public systems as long as proper precautions and necessary safety measures are in place \cite{Joselow2020}\cite{Sadik2020}\cite{SPF2020}. However, the public transit system may play some role in the spread of the disease \cite{CDC2020b} which is supported by the early cases in Hunan and Zhejiang, China \cite{Luo2020Hunan}\cite{Shen2020Zhejiang}, and the risks must be considered when a reopening plan is drafted for the transit systems. Transportation researchers are now trying to model the pandemic within public transportation systems \cite{Mo2021} \cite{Kumar2021} \cite{Liu2020}. Mo et al.\cite{Mo2021} modeled the pandemic using an public transportation encounter network (PTN) and an epidemic transmission model.

\subsection{Space-time transit network design problem}
\label{S:1.3}

The aim of this study is to integrate public transit network reopen plan and timetable optimization. The transit network reopen is a network design problem aiming to determine the network layout, frequencies, timetables, and ticket prices. Transportation network design has been studied for decades. The reader can refer to survey papers such as \cite{Farahani2013}\cite{Guihaire2008}\cite{Ibarra2015} and some of the latest papers on this topic \cite{An2016}\cite{Cancela2015}\cite{Fan2018}.

The integrated transit network and timetable design problem can be formulated as a two-level network design problem. The upper level involves the design problem with integer design variables, while the lower-level problem is the traffic assignment problem with path or link flow variables. The formulation of the lower-level problem depends on the assumption of the user routing choice. One typical assumption is user equilibrium (UE), that is, Wardrop's first principle. UE assumes that a Nash equilibrium will be achieved among all users once the network design is given. For example, in \cite{Gao2004}, the UE condition was formulated as variational inequalities. Their algorithm can determine a near-optimal network layout and frequency design of a transit system. However, the arc capacity constraints were not explicitly stated in their model. Verbas and Mahmassani\cite{Verbas2015} assumed an elasticity of travel demand and optimized the frequency allocation.

Another common assumption made when solving the integrated network and timetable design problem is the system optimal (SO), that is, Wardrop's second principle. It assumes that the manager can control the route choice of all users. When the SO is assumed, the two-level problem becomes a one-level optimization problem because the decision-maker determines all the decision variables, so the problem is usually easier to solve than UE. For example, in Niu's papers \cite{Niu2013} and \cite{Niu2015}, the authors only considered a railway corridor, and there were no transfers so that the SO could be defined analytically. Although they only considered one simple corridor, they used a space-time network with a time dimension, so their nonlinear network UE was still complicated to solve. Szeto and Jiang \cite{Szeto2014} considered a capacitated static transit network with transfers, and as the lower-level problem became linear, the SO could be solved efficiently.

Besides the SO and UE, users can be assumed to be boundedly rational, where the users will comply with the system assigned path as long as the path does not cost much more than the current shortest path. Liu and Zhou \cite{Liu2016} used this assumption and built an agent-based space-time network. Their model captured many subtle behaviors of the passengers, such as transfer and waiting time on platforms.

\subsection{Potential contributions and structure of this paper}
\label{S:1.4}

The contributions of this work can be summarized as follows: (1) Providing references to transportation managers. We will answer several important questions: How can the public transportation system look ahead to meet the challenge of demand during the pandemic? How can the existing transit network be reopened? How can we design an optimal schedule? (2) Novel formulation of a pandemic in public transit systems. We use a space-time dynamic transit network to capture complicated behaviors such as transfers and waiting at platforms, and we capture the risk of pandemic transmission using a simplified spatial compartmental model of epidemics (SCME) \cite{Birge2020} (3) Efficient algorithms for similar problems. Although the full formulation is difficult to solve because of a huge number of variables and constraints, we show that with Lagrangian relaxation, the problem can be decomposed into two subproblems. One subproblem involves a multi-commodity network flow problem, while an off-the-shelf solver such as Gurobi \cite{GUROBI2020} can be used to solve the other subproblem.

The remainder of this paper is organized as follows. In section \ref{S:2}, we present the assumptions of our model, and detail the formulation of our dynamic transit network considering the pandemic risk and social-distancing measures. A simple example is used to illustrate the major features of the dynamic network. In Section \ref{S:3}, we formulate the dynamic network design problem with pandemic risk as our objective function and social-distancing constraints, and the Lagrangian relaxation procedure used to solve this problem is presented in section \ref{S:4}. In section \ref{S:5}, we validate our model using real-world data from the BART system in the San Francisco Bay Area and provide an optimal timetable design. In section \ref{S:6}, the conclusions and potential future research questions are presented.

\section{Space-time transit network with epidemic/pandemic}
\label{S:2}
Consider a city planing to reopen public train/bus network under a pandemic similar to COVID-19. We will call this pandemic `the pandemic' in the following sections. We will build a space-time passenger flow network in section \ref{S:2.1}, and we will integrate epidemic/pandemic model with this network in section \ref{S:2.2}.

\subsection{Space-time transit network model}
\label{S:2.1}

\subsubsection{Physical network}
\label{S:2.1.1}
\begin{figure}[ht]
  \centering
  \includegraphics[width=0.8\linewidth]{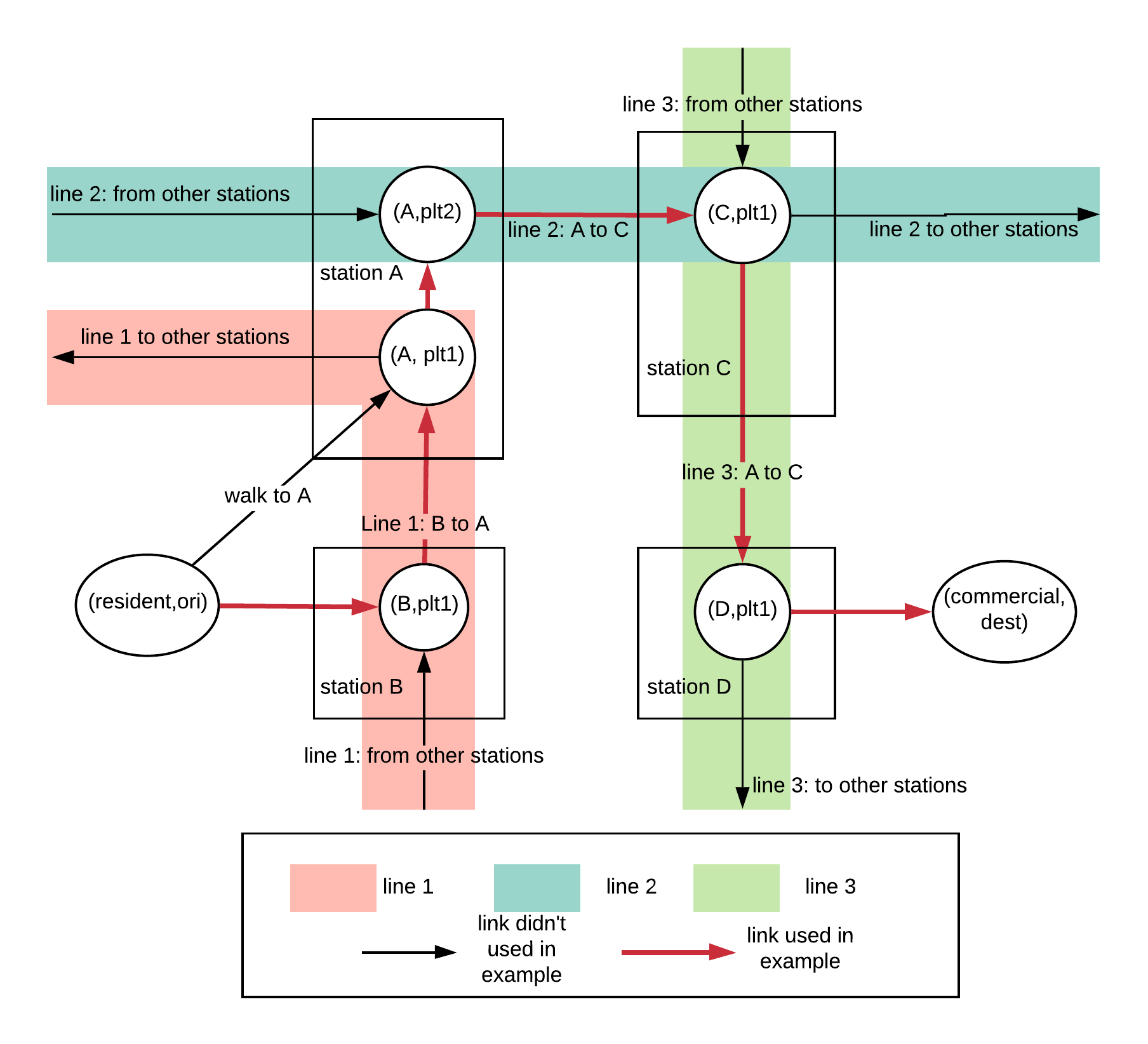}
  \caption{Physical network}
  \label{F:Physical}
\end{figure}

Consider a city with three types of locations: $\SSS \cup \OOO \cup \DDD$, where $\SSS$ is the set of stations, $\OOO$ is the set of origin areas, and $\DDD$ is the set of destination areas. In each station $s$, there could be one or many platforms $plt_j\in PLT(s)$, where $PLT(s)$ is the set of platforms in station $s$. In many real-world systems, multiple lines can share one platform while it is also common that one station has many platforms for different lines.

Let the set of all candidate bus/train lines be $\LLL=\{0,1,...,|\LLL|\}$, and the set of lines passing station $s\in \SSS$ using platform $plt_j$ be $\LLL(s,plt_j)$. One line consists several arcs connecting platforms from one station to another.

There are three types of nodes in our network: origin, destination, and platforms. We use tuple $(i,k)$ to index nodes in $\NNN^0$, where $i\in \SSS \cup \OOO \cup \DDD$ represent the physical location of the node, and $k \in \{ori,dest\}\cup \{plt_j\in PLT(s)\}$ represent the types of the nodes. $k$ can be any one of the following:

\begin{itemize}
    \item $ori$: origin
    \item $dest$: destination
    \item $plt_j$: platform $plt_j$ for in station $s$
\end{itemize}

For example, node $(A, plt_1)$ in figure \ref{F:Physical} is the platform 1 at station A, and node $(resident, ori)$ is the origin node at the `resident' area. 

Using this index rule, the set of nodes $\NNN^0$ is the union of set of origin nodes $\{(o,ori):o\in \OOO\}$, the set of destinations $\{(d,dest):d\in \DDD\}$, and the set of all platforms $\{(s,plt_j): s\in \SSS, plt_j \in PLT(s)\}$
\begin{align*}
    \NNN^0 :=& \{(o,ori):o\in \OOO\}\cup \{(d,dest):d\in \DDD\}\cup \{(s,plt_j): s\in \SSS, plt_j\in PLT(s)\}
\end{align*}

There are also several types of arcs, as is illustrated in figure \ref{F:Physical}, including walking arcs between origins/destinations and stations, in-vehicle travel arcs between platforms of the same line at two different stations, and transfer arcs from one platform to another at the same station. Note that there are two types of transfers: transfers using the same platform, and transfers using different platforms. We use set $\AAA^0$ to represent all arcs described above. The physical network is defined as
 \[ ( \NNN^0, \AAA^0) \]

We can illustrate the network using a simple example in figure \ref{F:Physical}. Given sets of physical locations $\OOO = \{resident,...\}$, $\DDD=\{commercial,...\}$, $\SSS=\{A,B,C,D...\}$ and set of lines $\LLL = \{1,2,3...\}$, where line 1 (red) connects station B to A, line 2 (blue) connects station A to C, and line 3 (green) connects station C to D. Consider one passenger travel from the `resident' area to the `commercial' area. He/she can take line 1 from station B to station A, transfers to line 2 at station A, transfers to line 3 at station C, takes line 3 from station A to station C, and walk to the commercial area. His/her trajectory is plotted in figure \ref{F:Physical} using red arcs. Note that in station A, there are two platforms so that transfer passengers need to change their platforms, while in station C, all lines use the same platform.

\subsubsection{Space-time network}
\label{st network}
During a pandemic like COVID-19, the capacity of each vehicle is strictly limited according to the social distancing rules. Therefore, we need to know the exact number of passengers on each vehicle and impose a strict social distancing capacity constraint, instead of just calculating the average number of passengers and compare it with the capacity. In addition, we need to track all passengers who stayed at the same vehicle or waited at the same platform with patients, since infection may happens on platforms or vehicles. Besides, we want to know the exact waiting time to optimize the timetable so that passengers can transfer between different lines smoothly.

We cannot answer these questions using the physical network in section \ref{S:2.1.1}, which can only give network flows averaged over time. We need to add an time dimension to the network and model the city as a space-time network similar to the models in \cite{Fan2018}\cite{Niu2013}\cite{Niu2015}. The idea is to make many copies of the nodes physical network and connect them using different types of arcs representing in-vehicle travelling, waiting on the platforms, walking, and early/late departure/arrival.

Recall that the set of locations is $\OOO \cup \SSS\cup \DDD$, the set of lines is $\LLL$, and for each station $s\in \SSS$, $\LLL(s,plt_j)\subset \LLL$ is the set of lines coming through station $s$ using platform $plt_j$. Consider a discrete time horizon $\TTT :=\{1,2,3,...\}$, we make $|\TTT|$ copies of $\NNN^0$ in the physical network and add a time index to each node. In this network, each node can be represented by a tuple $(i,k,t)$, where $i\in \OOO\cup\SSS\cup \DDD$ is the physical location, $k\in \{ori,dest\}\cup \{plt_j\in PLT(s)\}$ is the type of the node (defined in \ref{S:2.1.1}), and $t \in \TTT$ is the time step. Mathematically, the set of nodes is defined as: 
\[\NNN := \bigcup_{t\in \TTT} \NNN^t\]
Where we make $|\TTT|$ copies of the node set in physical network:
\begin{align*}
    \NNN^t :=& \{(o,ori,t):o\in \OOO\}\cup \{(d,dest,t):d\in \DDD\}\\
    & \cup \{(s,plt_j,t): s\in \SSS, plt_j\in PLT(s)\}
\end{align*}
In words, node set $\NNN^t$ at each time step $t$ is the union of the set of origin nodes, the set of destinations, and the set of all platforms. $\NNN$ is the union of all $|\TTT|$ of $\NNN^t$s.

Using the nodes defined, let the platform for line $l$ at station $s$ be $plt(l,s)$, we can construct the whole space-time network using the following 6 types of arcs:
\[\AAA:= \AAA_{transfer} \cup \AAA_{travel} \cup \AAA_{wait} \cup \AAA_{os} \cup \AAA_{sd} \cup \AAA_{dd}\]
Where sets are defined:
\begin{itemize}
    \item $\AAA_{transfer}:=\{(s,plt(l_1,s),t)\rightarrow (s,plt(l_2,s),t), \forall s\in \SSS, l_1,l_2\in \LLL(s), t\in \TTT\}$ contains arcs between two platforms the same station, where $\LLL(s)$ is the set of lines travelling through station $s$. Note that we ignore the walking time inside the stations.
    \item $\AAA_{travel}:=\{(s_1,plt(l,s_1),t)\rightarrow (s_2,plt(l,s_2),t+\Delta t_{s_1,s_2, l})|s_1,s_2\in \SSS, l\in \LLL(s_1)\cap\LLL(s_2)$, $t\in \TTT$, $s_1, s_2$ are two consecutive stations on line $l \}$ contains arcs between platforms of the same line at different stations. These are in-vehicle travel arcs, where $\Delta t_{s_1,s_2, l}$ is the train running time from $s_1$ to $s_2$ on line $l$, $plt(l,s)$ is the platform that line $l$ used at platform $s$.
    \item $\AAA_{wait}:=\{(s,plt(l,s),t)\rightarrow (s,plt(l,s),t+1)|s\in \SSS, l\in \LLL(s,plt_i)$, $t\in \TTT\}$ contains arcs between the same platform at different time period. These are platform waiting arcs where passengers can wait until the next available train/bus arrival, $plt(l,s)$ is the platform that line $l$ used at platform $s$.
    \item $\AAA_{sd}:=\{(s,plt_j,t)\rightarrow (d,dest,t+\Delta t_{s,d})|s\in \SSS,d\in \DDD, t\in \TTT, plt_j\in PLT(s)\}$ contains walking arcs from stations to destinations, where $\Delta t_{s,d}$  is walking time from $s$ to $d$.
    \item $\AAA_{os}:=\{(o,ori,t)\rightarrow (s,plt_j,t+\Delta t_{o,s})|s\in \SSS,o\in \OOO, t\in \TTT, plt_j\in PLT(s), \Delta_{min}<\Delta t_{o,s}<\Delta_{max}\}$ contains walking arcs from origins to stations, where $\Delta t_{o,s}$  is walking time from $o$ to $s$  plus early/late departure time difference. We allow passengers to leave earlier or later than their desired departure time, but the time difference is bounded: the time difference between desired departure time and the time that passenger enters one station must lies in $\Delta_{min},\Delta_{max}$. In our numerical examples, we set $\Delta_{min}=-30$ min and $\Delta_{max}=30$ min.
    \item $\AAA_{dd}:=\{(d,dest,t_1)\rightarrow(d,dest,t_2),\forall t_1,t_2\in \TTT\}$. These arcs connects all destination nodes of the same location together, so as long as the passenger arrive in their destination, they can complete their journey.
    
\end{itemize}
We define the set of time dependent OD pairs:
\[\WWW:=\{((o,ori,t),(d,dest,t))|o\in\OOO, d\in\DDD, t\in \TTT\}\]
and for each $w\in\WWW$, the demand is $d_w$.

\begin{figure}[ht]
\centering\includegraphics[width=1.\linewidth]{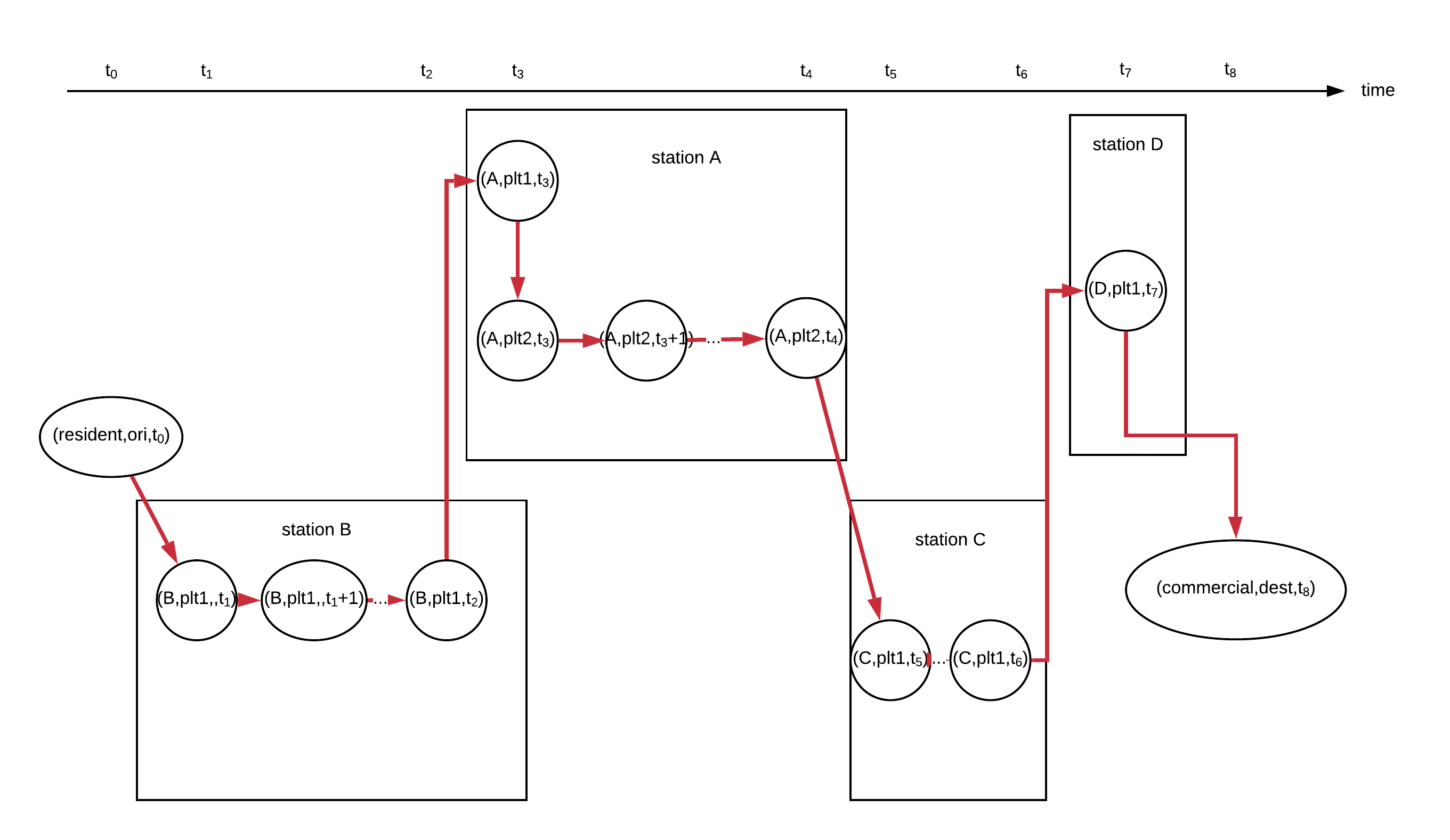}
\caption{Trajectory on space-time network}
\label{F:DN_F}
\end{figure}

Consider the previous example in figure \ref{F:Physical}, the passenger's trajectory on the space-time graph is in figure \ref{F:DN_F}. The whole process is:
\begin{enumerate}
    \item Walking from resident to station B's platform $plt_1$. This arc is in set $\AAA_{os}$.
    \[(resident, ori ,t_0)\rightarrow(B,plt_1,t_1)\]
    \item Waiting at the platform until the next train from line 1 arrives at station B. These arcs are in set $\AAA_{wait}$
    \[(B,plt_1,t_1)\rightarrow(B,plt_1,t_1+1)\rightarrow ...\rightarrow(B,plt_1,t_2)\]
    \item Riding a train/bus from B to A using line 1. $t_3-t_2$ is the in-vehicle travel time. This arc is from $\AAA_{travel}$
    \[(B,plt_1,t_2)\rightarrow (A,plt_1,t_3)\]
    \item Making transfer from line 1 to line 2 at station A, assuming the transfer takes $t_4-t_3$. This arc is in set  $\AAA_{transfer}$.
     \[(A,plt_1,t_3)\rightarrow (A,plt_2,t_3)\]
    \item Waiting at station A. These arcs are in set $\AAA_{wait}$.
    \[(A,plt_2,t_3)\rightarrow(A,plt_2,t_3+1)\rightarrow ...\rightarrow(A,plt_2,t_4)\]
    \item Riding line 2 to station C, assuming the travel time from station A to C on line 2 is $t_5-t_4$
    \[(A,plt_2,t_4)\rightarrow(C,plt_1,t_5)\]
    \item Waiting for line 3 at the same platform, and travelling to station D
    \[(C,plt_1,t_5)\rightarrow(C,plt_1,t_6)\rightarrow(D,plt_1,t_7)\]
    \item Walking to commercial area from station $C$, assuming it takes $t_8-t_7$  to walk from station C to 'commercial' area.
    \[(D,plt_1,t_7)\rightarrow(commercial, dest,t_8)\]
\end{enumerate}

\subsection{SCME model on the time-space network}
\label{S:2.2}
According to pandemic models like SCME \cite{Birge2020} models. We know that pandemic like COVID-19 transmission happens when susceptible passengers stay with patients in the same location for a certain period of time. In other words, a healthy person must spend some time with a patient together to get infected.

If we ignore the walking time inside stations (see assumption 4 in section \ref{S:3.1}), the pandemic transmission happens only when susceptible passengers stay with patients in the same vehicle or on the same platform for a period of time. Using the dynamics from SCME \cite{Birge2020}, we treat each origin node $o\in \OOO$ as the place where population lived, and each in-vehicle travel link and on-platform waiting link as place where transmission happens. We ignore the dead population since our time horizon is short. We consider people in three categories: healthy but not immune (susceptible), cured or vaccinated, and infected (exposed + clinical infected + subclinical infected). The daily new cases for population living in area $o\in \OOO$ is
\begin{equation} \label{eq: 1}
    \Delta IP_o =\beta \sum_{(i,j)\in \AAA_{travel}\cup \AAA_{wait}\cup \AAA_{transfer} }TT_{o,(i,j)} IR_{ij}
\end{equation}
where $IP_o$ is the infected population in area $o$, $\Delta IP_o$ is the expected number of new cases, $TT_{o,(i,j)}$ is the total time that susceptible people from area $o$ spent on link $(i,j)$, and $IR_{ij}$ is the infection rate of all passengers on link $(i,j)$. $\beta$ is the transmission rate ($\approx 1.12$ for COVID-19 \cite{Li2020}). For example, if $5$ susceptible passengers from origin $o$ stay in the same train from $i$ to $j$ and the travel time is $6$ minutes, then $TT_{o,(i,j)} = 5\times 6 = 30$ minutes. If there is one patients in a train from $i$ to $j$ with other 99 healthy (susceptible) passengers, then $IR_{ij} = 0.01$.

Let the network flow of $w\in \WWW$ on link $(i,j)$ be $u_{ij}^w$, the travel time of link $(i,j)$ be $c_{ij}$, and the ratio of susceptible population be $q_s$ (healthy people who have never been infected and have not taken vaccine) we can calculate the total time that people from $o$ spent on link $(i,j)$:
\begin{equation} \label{eq: 2}
    TT_{o,(i,j)} =\sum_{w:\ origin\ is \ o} u_{ij}^w c_{ij} q_s
\end{equation}
Let the probability that a person from OD pair $w$ is infected be $q_w$, we can calculate the infection rate on link $(i,j)$:
\begin{equation} \label{eq: 3}
    IR_{ij} =\frac{\sum_{w\in \WWW} u_{ij}^w q_w}{\sum_{w\in \WWW} u_{ij}^w}
\end{equation}
Plug \ref{eq: 2} and \ref{eq: 3} into \ref{eq: 1}, we can estimate the new infections in our transit system among people from origin $o$,
\begin{equation} \label{eq: 1full}
    \Delta IP_o =\beta q_s \sum_{(i,j)\in \AAA_{travel}\cup \AAA_{wait} \cup \AAA_{transfer}}
   \sum_{w_o:\ origin\ is \ o} u_{ij}^{w_o} c_{ij}  \frac{\sum_{w\in \WWW} u_{ij}^w q_w}{\sum_{w\in \WWW} u_{ij}^w}
\end{equation}
Therefore, we can simply sum all infections together to estimate the total number of new infections in our system:
\begin{equation} \label{eq: 5}
\begin{split}
    \Delta IP & =\beta q_s \sum_{(i,j)\in \AAA_{travel}\cup \AAA_{wait}\cup \AAA_{transfer} }
   \sum_{w_o\in \WWW} u_{ij}^{w_o} c_{ij}  \frac{\sum_{w\in \WWW} u_{ij}^w q_w}{\sum_{w\in \WWW} u_{ij}^w}\\
   & = \beta q_s \sum_{(i,j)\in \AAA_{travel}\cup \AAA_{wait}\cup \AAA_{transfer} }
   c_{ij} \sum_{w\in \WWW} u_{ij}^w q_w
\end{split}
\end{equation}

\section{Network and timetable design problem under the pandemic}
\label{S:3}
\subsection{Assumptions}
\label{S:3.1}
We make these key assumptions:
\begin{enumerate}
    \item Users are boundedly rational.
    \item Buses/trains are capacitated.
    \item The OD demand at each time step is deterministic, and the system will serve all the demand.
    \item Walking time inside stations is ignored.
    \item Bus/train travel time between two stations are fixed.
    \item Buses/trains need to be cleaned and disinfected at the last station after one run. 
    \item The train dwell time is ignored, and passenger boarding takes no time.
    \item Assumptions on the pandemic as in section \ref{S:2.2}.
\end{enumerate}

In assumption 1, we assume users are boundedly rational, similar to model in \cite{Liu2016}, that the system manager can control the flow of passengers to some degree, but only when the users are deviated from the shortest path by limited amount of time. The manager can control the flo by changing the ticket price, enforcing capacity rules so that passengers are forced to change route when a cart reaches its capacity, or releasing real-time on-board crowd data so that passengers know the risk and they will avoid crowded areas for their own safety concerns.

When demand is small and capacity constraints are loose, our optimization problem \ref{E: obj total} will give an all-pair shortest path solution, which is the only user equilibrium, because the objective function is linear. When demand is large, the optimization problem \ref{E: obj total pure} gives one of the equilibrium: one user cannot improve his/her cost by changing route, because all better routes are congested. With the boundedly rational assumption, we know this equilibrium is similar to the actual equilibrium in real life, where people greedily look for the shortest path. Moreover, during the pandemic, some transit agencies got temporary power to manage passenger flow network by rejecting entrance to stations. In this case, agencies can manage the network flow and system optimal/boundedly rational can be achieved. For example, in mid-March, the government requires that every passenger submits application before using the bus system \cite{Hubei2020}.

In assumption 2, we assume that bus/train is capacitated due to the social distancing order. Although in some public transportation system like SF BART, there is no supervisors in the stations and the limit-capacity social distancing rule seems not mandatory. However, BART provides real-time on-board crowd data to public \cite{BART2020b} and people who are concerned may check these information and decide whether boarding the next train or not. According BART's data, in most of the cases, the actual number of passengers are smaller than the requirement of social distancing capacity and the capacity constraints are valid.

In assumption 3, we assume that all time-dependent OD demand must be satisfied. Considering the fact that people who need public transportation system during a pandemic like COVID-19 usually don't have access to other means of transportation like private cars, while carpool service suspends and taxis are expensive. In addition, during a pandemic like COVID-19, people are required to `go out only when necessary' so travel demand are deemed `necessary'. Therefore, in our model, all demand must be satisfied, otherwise the problem is not feasible.

In assumption 4, we ignore the walking time inside the station so that we only need to consider the transmission on platforms and vehicles (see our SCME model in section \ref{S:2.2}). Noting that passengers spent most of their time waiting on platforms and staying in vehicles while using public transportation systems, and the time that passengers spend walking inside the stations (usually less than 1 min) can be ignored compared to waiting time and traveling time (can be $>10$ min), so this assumption is close to reality.

In assumption 5, we assume that the running time between two stations is constant. Because when train/bus follows the timetable, there is no congestion in most of the mass transit systems if no accident happens. The travel time from station to station can be deemed as fixed since there is little congestion for bus system during the pandemic, and the travel time is relatively stable in railway systems.
 
During the pandemic like COVID-19, we know trains/buses need to be disinfected after each run. We assume in assumption 6 that this happens only at the last station.

We also ignore the train/bus dwell time at stations as assumption 7, since vehicles in mass transit like BART spend most time travelling \cite{BART2020a}.

The last assumption is valid for public transportation systems: it is only possible that a healthy passenger to get the pandemic when an infected passenger stays within a certain distance.

\subsection{Notations}
We only list important notations used in our optimization model. You can find the definition of different sets of arcs and nodes in section \ref{S:2.2}.
\begin{longtable}{c c}
\caption{Notations}
    \label{tab:notation}\\
    \hline
    \multicolumn{2}{c}{Sets}\\
    \hline
         $\OOO$ & set of origin nodes (physical)\\
         $\DDD$ & set of destination nodes (physical)\\
         $\SSS$ & set of station nodes (physical)\\
         $\SSS_0$ & set of the first/last station for each line, $\SSS_0\subset \SSS$\\
         $\LLL$ & set of bus/train lines\\
         $\TTT$ & set of time\\
         $\NNN$ & set of all nodes in space-time network\\
         $\AAA$ &  set of all arcs in space-time network\\
         $\WWW$ & set of OD pairs\\
         $\LLL^{in}(s)$ & set of lines using $s$ as the last stop\\
         $\LLL^{out}(s)$ & set of lines using $s$ as the first stop\\
         $\PPP_w$ & set of all paths of OD pair $w$\\
         $\PPP_w'$ & set of paths of OD pair $w$ satisfying boundedly rational condition\\
    \hline
    \multicolumn{2}{c}{Continuous variables}\\
    \hline
     $\pi_{lt}$, $\mu_{s}$ & Lagrangian multipliers\\
        $f_w^p$ & flow on path $p$ in the path-flow formulation\\
        $u_{ij}^w$ & arc flow on $(i,j)$ of OD pair $w\in \WWW$\\
        $v_{ij}$ & arc flow on $(i,j)$\\
    \hline
    \multicolumn{2}{c}{Integer variables}\\
    \hline
        $x_{lt}$ & binary, $=1$ if line $l$ is opened and one train is dispatched at time $t$\\
        $y_{s}$ & binary, $=1$ if station $s\in \SSS$ is opened\\
        $z_l$ & binary, $=1$ if line $l\in \LLL$ is opened\\
        $M_s$ & number of trains available at station $s\in \SSS_0$ at the beginning of the day\\
    \hline
    \multicolumn{2}{c}{Parameters}\\
    \hline
        $\alpha_{ij}^p$ & $=1$ if path $p$ will use arc $(i,j)$\\
        $\beta$ & transmission rate \\
        $B$ & budget\\
        $c_{i,j}$ & link travel time of $(i,j)$\\
        $c_{i,j}^w$& link cost with multipliers of $(i,j)$ for users from $w$\\
        $C_p$& cost of path $w$ with multipliers\\
        $c_{l,open}$ & cost of opening a new line $l$\\
        $c_{l,close}$ & cost of closing an existing line $l$\\
        $c_{s,open}$ & cost of opening a new station $s$\\
        $c_{s,close}$ & cost of closing an existing station $s$\\
        $c_{l,t}$ & cost of dispatching one train at time $t$ of line $l$\\
        $d_w$ & demand of OD pair $w\in \WWW$\\
        $M_0$ & total number of available trains\\
        $N_{ij}$ & capacity of link $(i,j)$ under social distancing rules\\
        $q_w$ & infection rate of OD pair $w\in \WWW$\\
        $q_s$ & ratio of suseptiable people \\
        $z_l^0$ & $=1$ if line $l$ is already opened\\
        $SP(w)$& one of the shortest paths for OD pair $w$, $SP(w)\in \PPP_w$\\
        $T_l$ & time for line $l$ to complete a run from the first stop to the last stop\\
        $TL(w)$& tolerance limit for users from OD pair $w$ in boundedly rational condition\\
        $y_s^0$ & $=1$ if station $s$ is already opened\\
        \hline
\end{longtable}
\subsection{Budget constraints and agency cost}
For the manager, we need to consider the cost when a new line is opened or closed, which will cost
\[\sum_{l\in\LLL} c_{l,open} (z_l-z_l^0)^+ +c_{l,close}(z_l-z_l^0)^-\]
\[=\sum_{l\in\LLL} \bar{c}_l(z_l-z_l^0)\]
where\\
\begin{center}
    $\bar{c}_l = c_{l,open}$ if $z_l^0=0$\\
    $\bar{c}_l = -c_{l,close}$ if $z_l^0=1$\\
\end{center}
We also need to consider the cost of opening stations,
\[\sum_{s\in\SSS} c_{s,open} (y_s-y_s^0)^+ +c_{s,close}(y_s-y_s^0)^-\]
\[=\sum_{s\in\SSS} \bar{c}_s (y_s-y_s^0)\]
where
\begin{center}
    $\bar{c}_s = c_{s,open}$ if $y_s^0=0$\\
    $\bar{c}_s = -c_{s,close}$ if $y_s^0=1$\\
\end{center}
For the cost of dispatching one train run, we have
\[\sum_{t\in\TTT} \sum_{l\in \LLL} c_{lt}x_{lt}\]
The cost of dinsinfection is included in this cost.

Sum all these costs together and consider our budeget, we have the following budget constraint:
\begin{equation}\label{E: budget}
    \sum_{t\in\TTT} \sum_{l\in \LLL} c_{lt}x_{lt}
    +\sum_{l\in\LLL} \bar{c}_l(z_l-z_l^0)
    +\sum_{s\in\SSS} \bar{c}_s (y_s-y_s^0) \leq B
\end{equation}

\subsection{Fleet design constraints}
Using the definition from table \ref{tab:notation}, we define binary variables $x_{lt}=1$ if we run one train on line $l$ at time $t$ from its first station, $z_l=1$ if line $l$ is opened, and $y_s=1$ if station $s$ is opened. We know the binary design variables must comply:

\begin{equation}\label{E:xr}
    x_{lt}\leq z_l, \forall l\in \LLL, t\in \TTT
\end{equation}
\begin{equation}\label{E:binary constr}
   x_{lt}, z_l, y_s \in \{0,1\}, \forall l,t,s
\end{equation}
In addition, we must have enough bus at each first/last station through the time horizon.
\begin{equation}\label{E: bus number nonnegative}
    M_s -\sum_{l\in\LLL^{out}(s)}\sum_{t\leq t_1} x_{lt}
    + \sum_{l\in\LLL^{in}(s)}\sum_{t\leq t_1-T_l} x_{lt}
    \geq 0, \forall t_1\in \TTT, \forall s\in \SSS_0
\end{equation}
\begin{equation}\label{E: total bus number}
    \sum_{s\in \SSS_0} M_s\leq M_0
\end{equation}
In \ref{E: bus number nonnegative}, the LHS is the total number of vehicles at time $t_1$ at the station $s$, and we need to make sure that there are always non-negative number of vehicles at each station. We define $\LLL^{in}(s)$ the set of lines using station $s$ as the last station, $\LLL^{out}(s)$ the set of lines using $s$ as the first station, and $T_l$ is the time for $l$ to complete a run (from the first to the last station). In \ref{E: total bus number}, the total number of vehicles at time $0$ must be less or equal than the total available vehicle number.

\subsection{Network flow constraints}
We use the path-flow formulation in our model, let $\PPP_w$ be the set of paths for OD pair $w$, and let $f_w^p$ be the flow on path $w$ for OD pair $w$. Since we consider deterministic OD demand, we have the network flow conservation:
\begin{equation}\label{E:flow conservation}
   \sum_{p\in \PPP_w} f_w^p = d_w, \forall w\in \WWW
\end{equation}and
\begin{equation}\label{E:flow nonnegative}
     f_w^p \geq 0, \forall w\in\WWW,p\in\PPP_w
\end{equation}
We can calculate the flow on each arc using path flow, using the 0-1 indicator $\alpha^p_{ij}$: $\alpha^p_{ij}=1$ if path $p$ takes arc $(i,j)$, $=0$ otherwise.
\begin{equation}\label{E:OD flow}
    u_{ij}^w = \sum_{p\in\PPP_w} \alpha_{ij}^p f_w^p,\forall (i,j)\in \AAA
\end{equation}
\begin{equation}\label{E:flow total}
    v_{ij} = \sum_{w\in\WWW}\sum_{p\in\PPP_w} \alpha_{ij}^p f_w^p,\forall (i,j)\in \AAA
\end{equation}
We also have the social-distancing capacity constraints for each vehicles:
\begin{equation}\label{E:flow cap}
    v_{ij}\leq N_{ij}, \forall (i,j)\in \AAA
\end{equation}
Consider the design variables, no flows are allowed if there is no available line/station/train run, which are:
\begin{equation}\label{E:flow cap design 1}
\sum_{(i,j)\in \AAA_{travel}(l,t)} v_{ij}\leq x_{lt} \sum_{(i,j)\in \AAA_{travel}(l,t)}N_{ij} , \forall l\in \LLL, t\in \TTT
\end{equation}
where $\AAA_{travel}(l,t)\subset\AAA_{travel}$ is the set of arcs corresponding to train/bus of line $l$ departure at time $t$
\begin{equation}\label{E:flow cap design 2}
\sum_{(i,j)\in \AAA_{os}(s)\cup \AAA_{sd}(s)}v_{ij}\leq  y_s\sum_{(i,j)\in \AAA_{os}(s)\cup \AAA_{sd}(s)} N_{ij}, \forall s\in \SSS
\end{equation}
where $\AAA_{os}(s)\subset\AAA_{os}$, $\AAA_{sd}(s)\subset\AAA_{sd}$ are the sets of arcs correspond to station $s$.

\subsection{Boundly rational constraints}
For each OD pair $w$, there would be a shortest path solution $SP(w)$ on physical network, the travel cost of this path is $\sum_{(i,j)\in SP(w)} c_{ij}$.
We have the boundedly rational constraint:
\begin{equation}\label{E:boundedly rational}
(\sum_{(i,j)\in p} c_{ij} -\sum_{(i,j)\in SP(w)} c_{ij}-TL(w))f_w^p \leq 0, \forall p\in \PPP_w, w\in \WWW
\end{equation}
This constraint means that the route that the users choose cannot be much worse from the shortest path. This set of constraints can also be written as $p\in \PPP_w'$, where $\PPP_w'$ is the set of paths satisfying boundedly rational conditions.

\subsection{Objective function and the optimization problem}
\label{S:3.7}
We are looking forward to control the total number of pandemic contraction within the system while all passengers are satisfied. Using the expression of total new infections from equation \ref{eq: 5} in section \ref{S:2.2} , the problem is
\begin{equation}\label{E: obj total pure}
    \min_{u,v,x,y,z,r,M_s} {
   \beta q_s \sum_{(i,j)\in \AAA_{travel}\cup \AAA_{wait}}
   c_{ij} \sum_{w\in \WWW} u_{ij}^w q_w
    }
\end{equation}
such that 
\ref{E: budget}, \ref{E:xr}, \ref{E:binary constr}, \ref{E: bus number nonnegative}, \ref{E: total bus number}, \ref{E:flow conservation}, \ref{E:flow nonnegative},\ref{E:OD flow}, \ref{E:flow total}, \ref{E:flow cap}, \ref{E:flow cap design 1}, \ref{E:flow cap design 2}, and \ref{E:boundedly rational} are satisfied.

Note that the goal to minimize the pandemic contraction coincide with the goal to minimize total passenger travel time since we have a linear objective function with positive coefficient proportional to link travel time.

To make the problem feasible when doing numerical iterations. We add arcs $\AAA_{od}:=\{(o,ori,t)\rightarrow(d,dest,t), o\in \OOO, d\in \DDD, t\in \TTT\}$ between origin and destination nodes with large penalty $c_{ij}=M$, and flow on this path doesn't subject to rational bounded condition \ref{E:boundedly rational}. The problem that we solved in numerical examples are:

\begin{equation}\label{E: obj total}
    \min_{u,v,x,y,z,r,M_s} {
   \beta q_s \sum_{(i,j)\in \AAA_{travel}\cup \AAA_{wait}}
   c_{ij} \sum_{w\in \WWW} u_{ij}^w q_w +\sum_{(i,j)\in \AAA_{od}} c_{ij}v_{ij}
    }
\end{equation}

Here the second term is not the travel time, but a large penalty for unsatisfied demand. Arcs in $\AAA_{od}$ can be used only when this OD demand cannot be served and problem is infeasible. We cannot directly add the travel cost and the cost of new cases since the travel cost is in short term while the cost of new infection is in long term, and it's difficult to estimate the net present value of pandemic cost, or the value of time for each passenger during the pandemic. Also, we know that minimizing the total pandemic contraction can help minimizing total travel time, therefore, We don't need to add link travel cost to the objective function.

\section{Solving the problem using Lagrangian relaxation}
\label{S:4}
\subsection{Problem relaxation and decomposition}
To find the optimal design, we can use Lagrangian relaxation method by relaxing constraints \ref{E:flow cap design 1} and \ref{E:flow cap design 2}. Let multipliers of \ref{E:flow cap design 1} be $\pi_{lt}$ for each line $l$ at time $t$, and multipliers for \ref{E:flow cap design 2} be $\mu_{s}$ for station $s$. We have the following dual function:
\begin{multline}
    L(\pi,\mu) := \min_{u,v,x,y,z,r,M_s}
    \beta q_s \sum_{(i,j)\in \AAA_{travel}\cup \AAA_{wait}}
   c_{ij} \sum_{w\in \WWW} u_{ij}^w q_w + \sum_{(i,j)\in\AAA_{od}} c_{ij}v_{ij} \\
   +\sum_{t\in \TTT, l\in \LLL}\pi_{lt}(\sum_{(i,j)\in \AAA_{travel}(l,t)} v_{ij}-x_{lt} \sum_{(i,j)\in \AAA_{travel}(l,t)}N_{ij})\\
   +\sum_{s\in\SSS}\mu_s(\sum_{(i,j)\in \AAA_{os}(s)\cup \AAA_{sd}(s)}v_{ij}-  y_s\sum_{(i,j)\in \AAA_{os}(s)\cup \AAA_{sd}(s)} N_{ij})
\end{multline}
such that \ref{E: budget}, \ref{E:xr}, \ref{E:binary constr}, \ref{E: bus number nonnegative}, \ref{E: total bus number}, \ref{E:flow conservation}, \ref{E:flow nonnegative}, \ref{E:OD flow}, \ref{E:flow total}, \ref{E:flow cap} and \ref{E:boundedly rational} are satisfied.

Therefore, we can decompose the problem into two sub problems,
\begin{equation}
L(\pi,\mu)= SUB_1 (\pi,\mu) + SUB_2 (\pi, \mu)
\end{equation}
The first sub problem is:
\begin{equation} \label{sub1}
\begin{split}
    SUB_1(\pi,\mu) &= \min_{u,v}
   \beta q_s \sum_{(i,j)\in \AAA_{travel}\cup \AAA_{wait} }
   c_{ij} \sum_{w\in \WWW} u_{ij}^w q_w \\
   &+ \sum_{(i,j)\in\AAA_{od}} c_{ij}v_{ij}
   +\sum_{(i,j)\in\AAA_{travel}} \pi_{lt(i,j)}v_{ij} 
   +\sum_{(i,j)\in\AAA_{os}\cup\AAA_{sd}} \mu_{s(i,j)}v_{ij}
\end{split}
\end{equation}
such that \ref{E:flow conservation}, \ref{E:flow nonnegative}, \ref{E:OD flow}, \ref{E:flow total}, \ref{E:flow cap}, and \ref{E:boundedly rational}

where $lt(i,j)$ is the line and time correspond to arc $(i,j)$ if $(i,j)\in\AAA_{travel}$, and $s(i,j)$ is the station that correspond to arc $(i,j)$ if $(i,j)\in\AAA_{os}\cup\AAA_{sd}$.

The second sub problem is
\begin{multline}
    SUB_2(\pi,\mu) = \min_{x,y,z,r,M_s}
    -\sum_{l\in \LLL, t\in \TTT }( \pi_{lt} x_{lt} \sum_{(i,j)\in \AAA_{travel}(l,t)}N_{ij} x_{lt})\\
   -\sum_{s\in \SSS} (\mu_{s}y_s \sum_{(i,j)\in \in\AAA_{os}(s)\cup\AAA_{sd}(d)}N_{ij})
\end{multline}
such that \ref{E: budget}, \ref{E:xr}, \ref{E:binary constr}, \ref{E: bus number nonnegative}, \ref{E: total bus number}.

\subsection{Solving the subproblems}
For $SUB_2$, we have an integer programming with small number of integer variables, which can be solved efficiently using solvers like GUROBI \cite{GUROBI2020}. If the manager don't care about the fleet problem, constraints  \ref{E: bus number nonnegative}, \ref{E: total bus number} can be ignored and this problem become a bin packing problem.

$SUB_1$ is a linear min-cost multi-commodity network flow problem with boundedly rational route choice conditions. We have introduced the path flow variable in the last section, here we can rewrite all link flow variable using path flow variables, and then solve it using column generation. We can reformulate $SUB_1$ as
\begin{equation}\label{sub1'}
    SUB_1':=\min_f \sum_{w\in\WWW}\sum_{p\in\PPP_w}\sum_{(i,j)\in p} c_{ij}^{w} f_w^p
\end{equation}
such that \ref{E:flow conservation}, \ref{E:flow nonnegative}, \ref{E:boundedly rational} and
\begin{equation}\label{E: cap path}
    \sum_{w\in\WWW}\sum_{p\in\PPP_w} \alpha_{ij}^p f_w^p\leq N_{ij},\forall (i,j)\in \AAA
\end{equation}
where $c_{ij}^{w}=\beta q_s c_{ij} q_w +\pi_{lt(i,j)}$ for all $(i,j)\in \AAA_{travel}$, $c_{ij}^w=\beta q_s c_{ij} q_w$ for all $(i,j)\in \AAA_{wait}$, $c_{ij}^{w}=\mu_{s(i,j)}\forall (i,j) \in \AAA_{os}\cup \AAA_{sd}$, $c_{ij}^w=c_{ij}$ if $(i,j)\in \AAA_{od}$, and $c_{ij}^w=0$ for all other arcs. We call $c_{ij}^w$ the link cost with multipliers. Note that \ref{E: cap path} is equivalent to \ref{E:flow cap} and \ref{E:flow total} combined.

Note that although $SUB_1'$ has exponentially many variables, the number of constraints are significantly smaller. We can apply column generation techniques to $SUB_1'$ with \ref{E:flow conservation}, \ref{E:flow nonnegative}, \ref{E: cap path} and \ref{E:boundedly rational}:

Given a set of paths, if the dual variables for each equation in flow conservation \ref{E:flow conservation} is $\lambda_w$, and the dual variables for each capacity constraint in \ref{E: cap path} is $\lambda_{ij}$, then the reduced cost for each path flow $f_w^p$ is $C_p-\lambda_w-\sum_{ij\in p} \lambda_{ij}$, where we define the total path cost
\[C_p = \sum_{(i,j)\in p} c_{ij}^w\]
To find the minimum reduced cost for one OD pair $w$, define all the set:
\begin{equation}
    \PPP_w':=\{p\in \PPP_w| p\ satsisfies \ \ref{E:boundedly rational}\} \cup \{p=(w_o,w_d)\}
\end{equation}
Here, $\PPP_w'$ contains all paths for OD pair $w$ that satisfies boundedly rational condition and one path that directly connect origin to destination but with large link cost as penalty. We only need to solve the following problems to generate a path: 
\begin{equation}\label{E:reduced cost}
    -\lambda_w + \min_{p\in\PPP_w'} (C_p-\sum_{ij\in p} \lambda_{ij})
\end{equation}
If $-\lambda_w + \min_{p\in\PPP_w'} (C_p-\sum_{ij\in p} \lambda_{ij})<0$, we can generate a new path using the path with minimum $(C_p-\sum_{ij\in p} \lambda_{ij})$. Since $\lambda_w$ is a constant given $w$, $C_p> 0$ and $\lambda_{ij} \leq 0$, this path can be found by solving a shortest path problem with non-negative link cost. If the minimum reduced cost is negative, we can add this path to basis and do one simplex iteration. Note that we only search paths in $\PPP_w'$, if the shortest path (with respect to link cost $c_{ij}^w$) that we found doesn't satisfy \ref{E:boundedly rational}, and the reduced cost of this path is negative, we will try to find the second shortest path and so on, until the reduced cost in \ref{E:reduced cost} become non-negative.

\subsection{Updating multipliers}
The upper bound can be found using the opening plan from $SUB_2$. We then apply the column generation algorithm on this network (with opened link only, and multipliers $\pi=0$ and $\mu=0$) to find the network flow solution. Note that this problem is always feasible using objective function in \ref{E: obj total}, but the cost will be extremely large if some demand cannot be satisfied by the public transportation system.

We know the dual function is concave, the subgradient for each $\pi_{lt}$ is $SG_{lt}:=\sum_{(i,j)\in \AAA_{travel}(l,t)} v_{ij}-x_{lt} \sum_{(i,j)\in \AAA_{travel}(l,t)}N_{ij}$, and the subgradient for each $\mu_{s}$ is $SG_{s}:=\sum_{(i,j)\in \AAA_{os}(s)\cup \AAA_{sd}(s)}v_{ij}-  y_s\sum_{(i,j)\in \AAA_{os}(s)\cup \AAA_{sd}(s)} N_{ij}$.
Our Lagrangian relaxation (LR) algorithm is summarized below:

\noindent\textbf{LR Algorithm}
\begin{enumerate} [{\textbf{Step }}1{. }]
    \item Initialize multipliers, and solve $SUB_1$ with these multipliers using column generation.
    \item Solve $SUB_2$ using solvers
    \item Check if $|SUB_2-SUB_1|$ is small enough or we have done more than 1000 iterations.
    \item Find an upper bound based on the result from $SUB_2$ using column generation algorithm for $SUB_1$, but only with opened links and 0 multipliers.
    \item Calculate sub-gradient (i.e. constraint violation), update the multipliers and integer variables using \textbf{Line search}. Go to step 3.
\end{enumerate}
The step length is chosen using the equation below and line search,
\begin{equation}\label{steplength}
    \gamma^t=\alpha^t \frac{UB-LB(\pi^t,\mu^t)}{\sum_{l,t}(SG_{lt})^2+\sum_s(SG_s)^2}
\end{equation}
where $UB$ is the best upper bound obtained so far and $LB(\pi^t,\mu^t)$ is the value of dual function at this iteration. We set $\alpha^t=0.1$ at first. At each LR iteration, we will do several steps of line search

\noindent\textbf{Line search}
\begin{enumerate} [{\textbf{Step }}1{. }]
    \item Calculate $\gamma^t$ using \ref{steplength}, $k=1$
    \item Calculate $UB'=UB((\pi_{lt}+\gamma^t SG_{lt})^+, (\mu_s+\gamma^t SG_s)^+)$ by solving $SUB_1((\pi_{lt}+\gamma^t SG_{lt})^+, (\mu_s+\gamma^t SG_s)^+)$ using column generation and $SUB_2((\pi_{lt}+\gamma^t SG_{lt})^+, (\mu_s+\gamma^t SG_s)^+)$ using MIP solver.
    \item If $UB'<UB$ and $k<max_k$, then $k=k+1, \gamma^t = \gamma^t/2$ and go to step 2, otherwise update $\pi_{lt}=(\pi_{lt}+\gamma^t SG_{lt})^+,\mu_s= (\mu_s+\gamma^t SG_s)^+$ and stop.
\end{enumerate}

\section{Numerical examples}
\label{S:5}

\subsection{A toy example}
\label{S:num toy}
\subsubsection{Parameters}
Here, we evaluate the current open plan and optimize the network design. The following parameters need to be estimated:
\begin{itemize}
    \item Fixed cost $c_l$, $c_t$
    \item Operation cost for one run $c_{lt}$
    \item Train and platform capacity under social distancing rules $N_{ij}$
    \item Pandemic infection rate at different locations $q_w$ and proportion of susceptible population $q_s$. 
    \item Transmission rate of the pandemic $\beta$
\end{itemize}

In this model, the contingency of the pandemic represented by $\beta$. According to \cite{Li2020}, $\beta = 1.12 person/day = \frac{1.12}{60\times24} person/min$. The definition of this parameter can be shown using an example: if we have $1000$ susceptible people spent one day in a place where 5\% of population is infected, then the expected number of new infection is $1000\times0.05\times1.12=56$ persons. We set the time tolerance for each user to be $TL=30$ min in the toy model.

The cost parameters estimated in the BART example in section \ref{S: bart} are used. In addition, the pandemic infection rates in different counties are extracted using the JHU data \cite{JHU} on May 10, 2021. We assume that stations in this toy problem are located in different counties in SF bay area. We set the cost of opening one line to be twice the price of operating one train run.

The OD demand is generated randomly from a simulated Poisson distribution. At every minute, for every OD pair, the demand is $Poisson(Demand)$ distributed, where $Demand$ is the demand intensity. We set $Demand=5/6$ in this example ($50$ passengers per hour).

\subsubsection{Validation of algorithm}
\begin{figure}[h]
    \centering
    \includegraphics{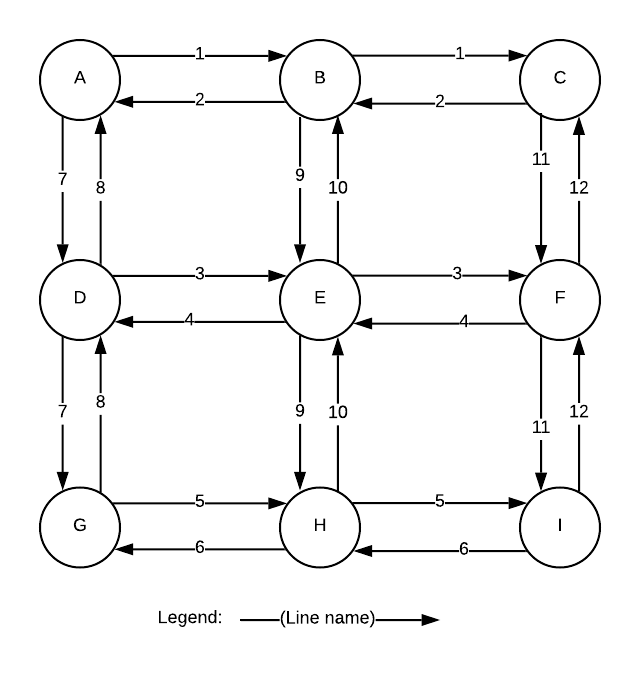}
    \caption{Toy train example}
    \label{fig:toy map}
\end{figure}

Consider a $3\times 3$ grid city network in figure \ref{fig:toy map}. We have 9 physical nodes $\OOO = \DDD = \SSS = \{A,B,...,I\}$ and 12 lines in our candidate set $\LLL =\{1,2,...,12\}$. Let in-vehicle travel time between two station be 5 min and walking time be 25 min. Let the train capacity be 600 and the total number of available trains be 50.

\begin{figure}[h]
    \centering
    \includegraphics[width=0.8\linewidth]{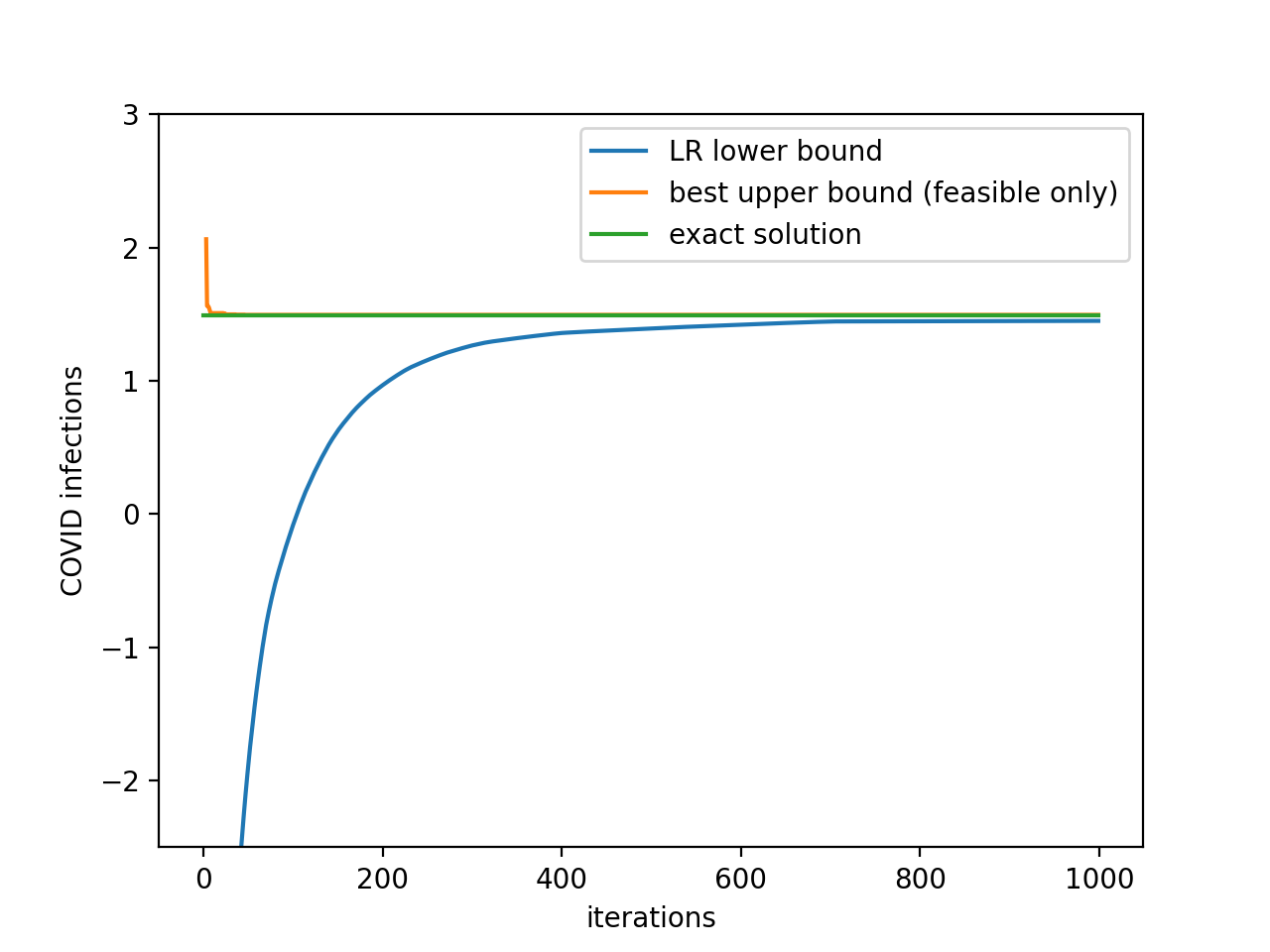}
    \caption{LR iteration of toy example}
    \label{fig:toy iter}
\end{figure}

As this is an easy problem, it can be solved using GUROBI solver \cite{GUROBI2020} and compared with the lower bound determined using the Lagrangian relaxation. The iteration process is shown in figure \ref{fig:toy iter}. The lower/upper bound after 1000 iterations is (1.4497, 1.4947), while the exact solution of the solver is 1.4947 in the toy problem. The proposed algorithm has a relatively tight bound ($<4\%$) for the toy problem, and the optimal solution is found when solving the upper bound. Therefore, our algorithm can correctly identify the optimal solution in this example.

\subsubsection{Sensitivity analysis on budget}
\begin{figure}[h]
\centering
\subfloat[\centering COVID cases]{
{\includegraphics[width=0.4\linewidth]{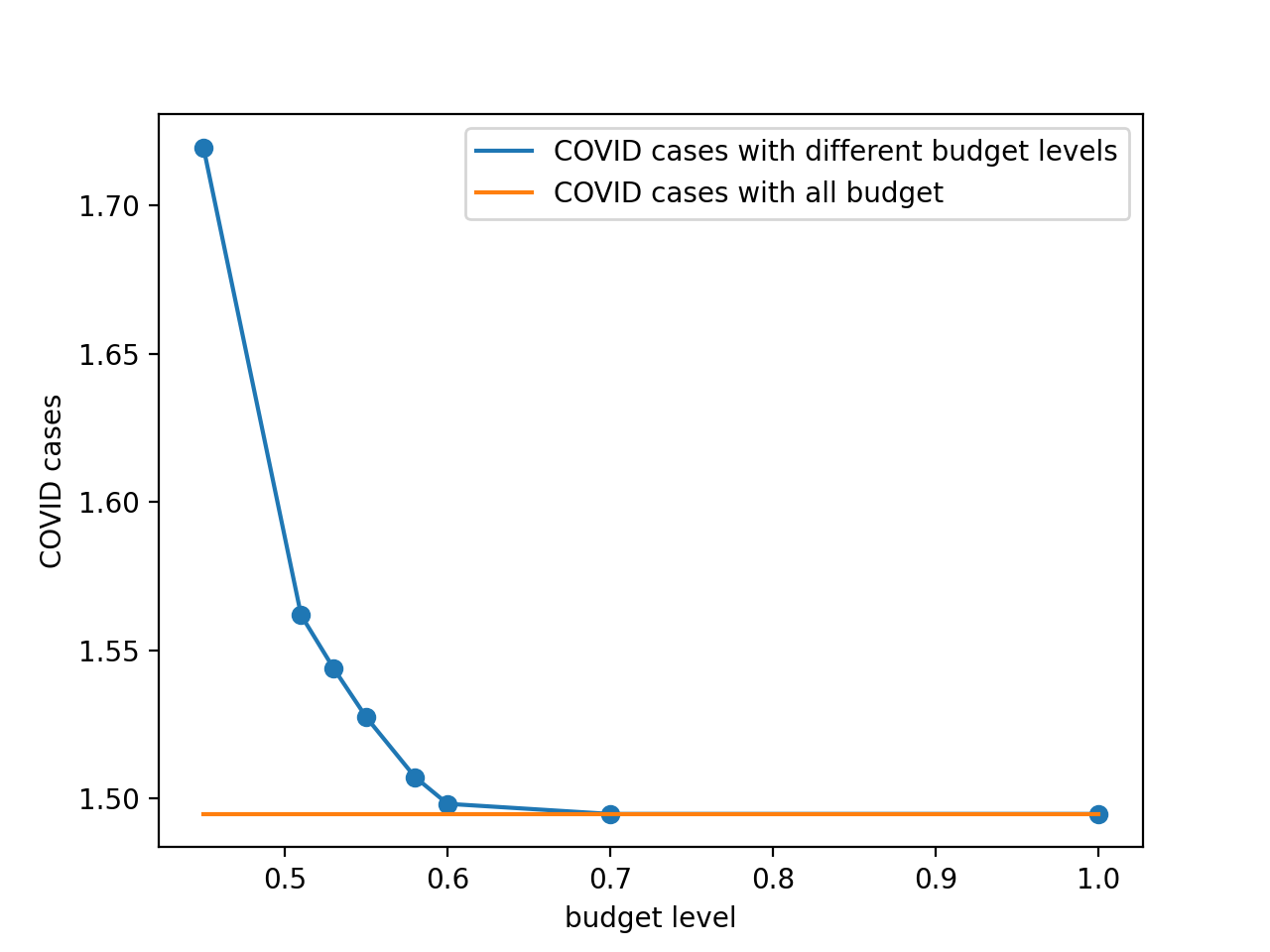} }}
\qquad
\subfloat[\centering total travel time]{
{\includegraphics[width=0.4\linewidth]{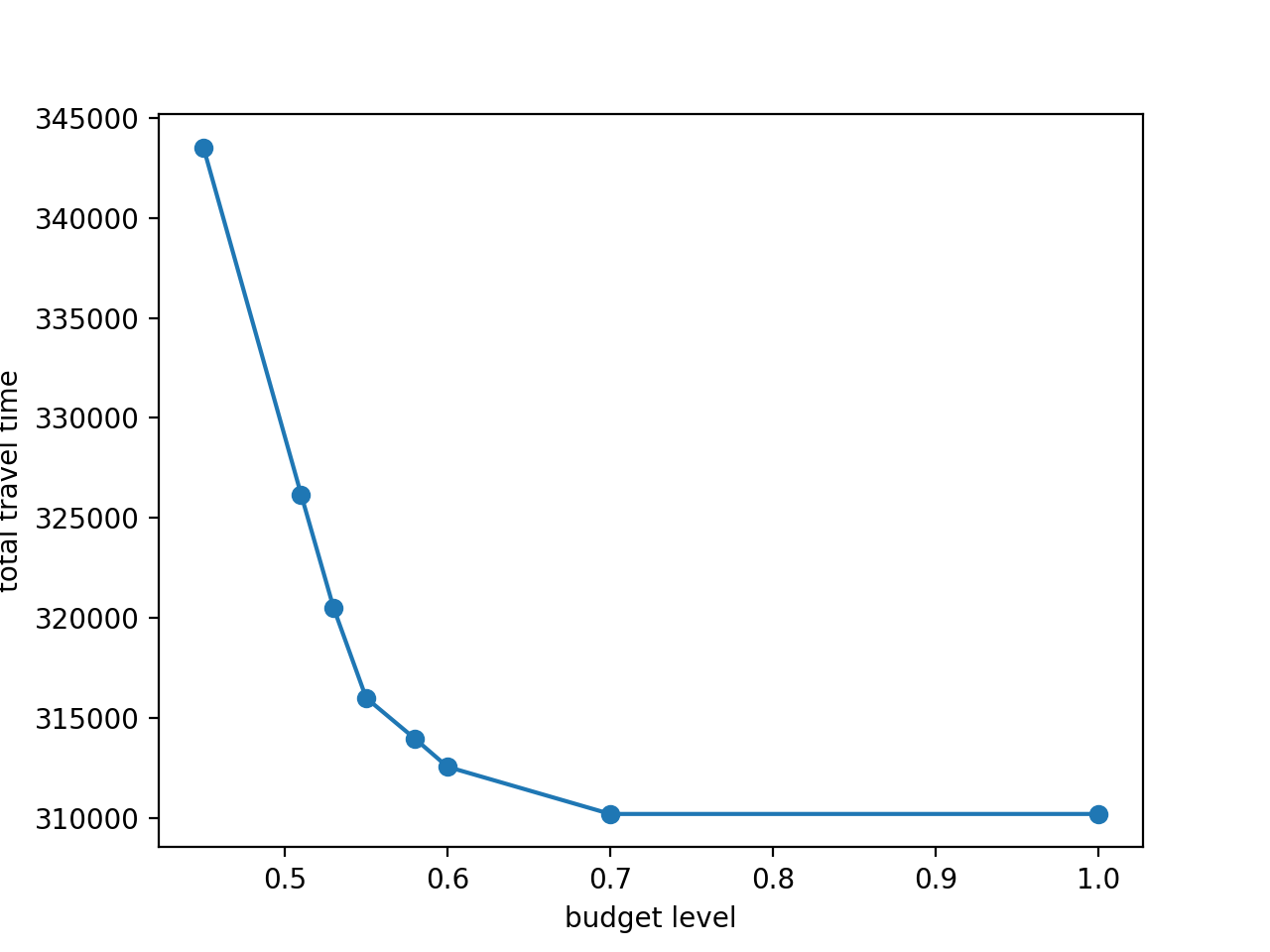} }}
\caption{Sensitivity analysis on budget}
\label{fig:toy budget}
\end{figure}

When budget is limited, we cannot operate too many train/bus runs. The total number of pandemic infection will increase as we cut the budget (See \ref{fig:toy budget}). We can see there is a minimum budget where the model become infeasible, and both of the pandemic cases and the total travel time are more sensitive as the budget level gets closer to this minimum budget (transition point). 

To show the impact of budget, we will use the best upper bound result after 1000 LR iterations for different parameters. We set the capacity to be 600. Note that our algorithm is just a heuristic and might not be the actual optimal solution, but just an upper bound.

In this example, when budget is greater than 70\% of the total budget provided in the last example, we can see the optimal solution will not change since the budget constrain is loose. When budget is smaller than 70\%, the pandemic infection (our objective function) will increase when we cut the budget further. When budget is smaller than 45\% of the total budget, the problem could be infeasible, since we cannot find a feasible solution satisfying all demand after 1000 LR iterations. Therefore, the budget transition point is $<45\%$.

Note that the pandemic cases (objective function) is more sensitive when budget level gets smaller (see fig \ref{fig:toy budget} (a)). This is because the LP relaxed lower bound
\begin{center}
    $F(b):=$ LP relaxed optimal objective function with budget $b$
\end{center}
 is convex with respect to $b$. Although the actual optimal objective function is not convex with respect to budget since we have a lot of integer variables, the trend is similar: the new pandemic cases is not sensitive when budget is great, but will be sensitive when budget is small.
 
 For total travel time, we can see in fig \ref{fig:toy budget} (b) that the trend of total travel time is the same as the pandemic cases, even if we don't have total travel time explicitly in our objective function. This is because the goal to minimize the pandemic contraction coincide with the goal to minimize total passenger travel time (see section \ref{S:3.7}). 

\subsubsection{Sensitivity analysis on capacity}
\begin{figure}[h]
\centering
\subfloat[\centering COVID cases]{
{\includegraphics[width=0.4\linewidth]{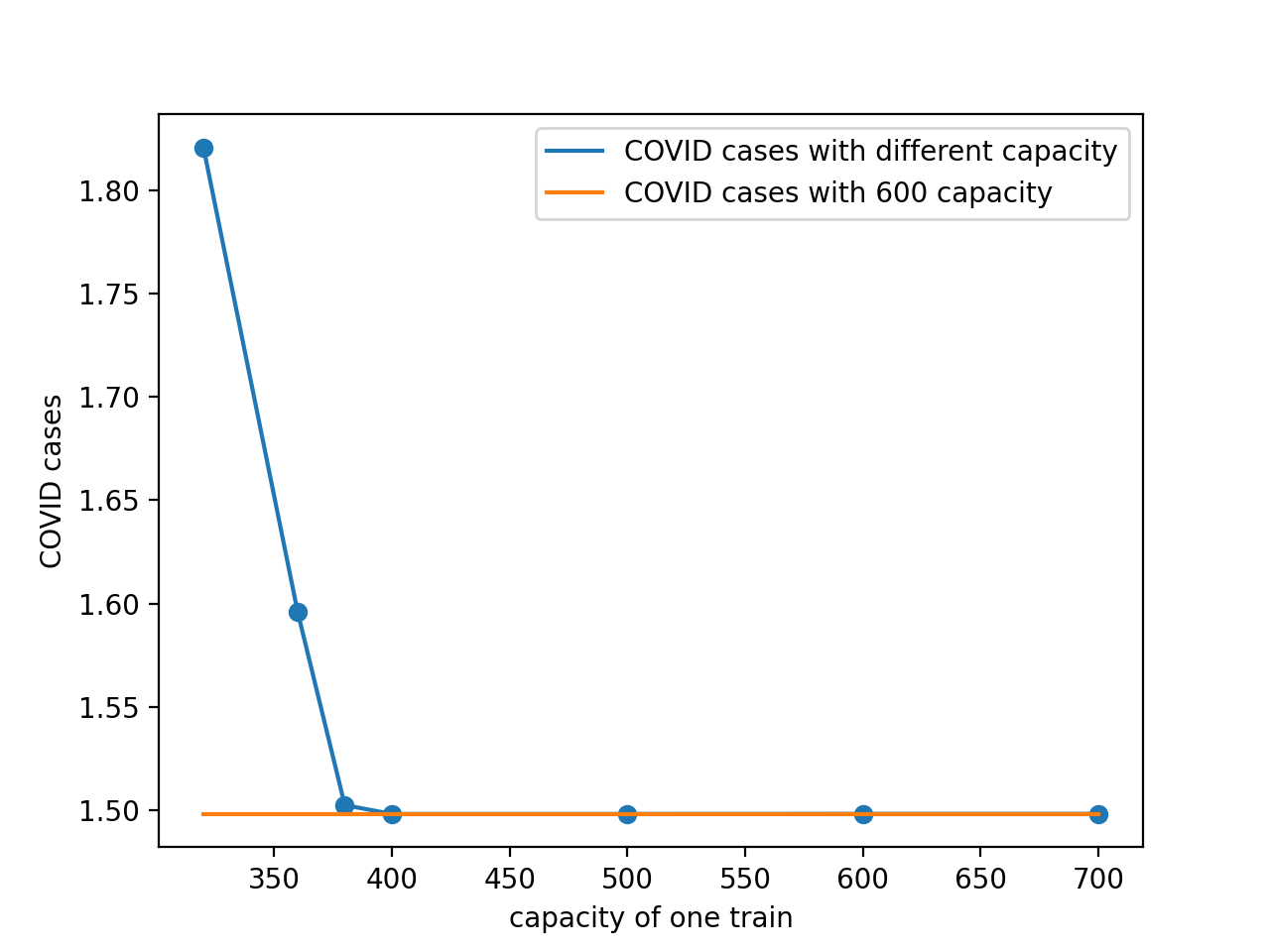} }}
\qquad
\subfloat[\centering total travel time]{
{\includegraphics[width=0.4\linewidth]{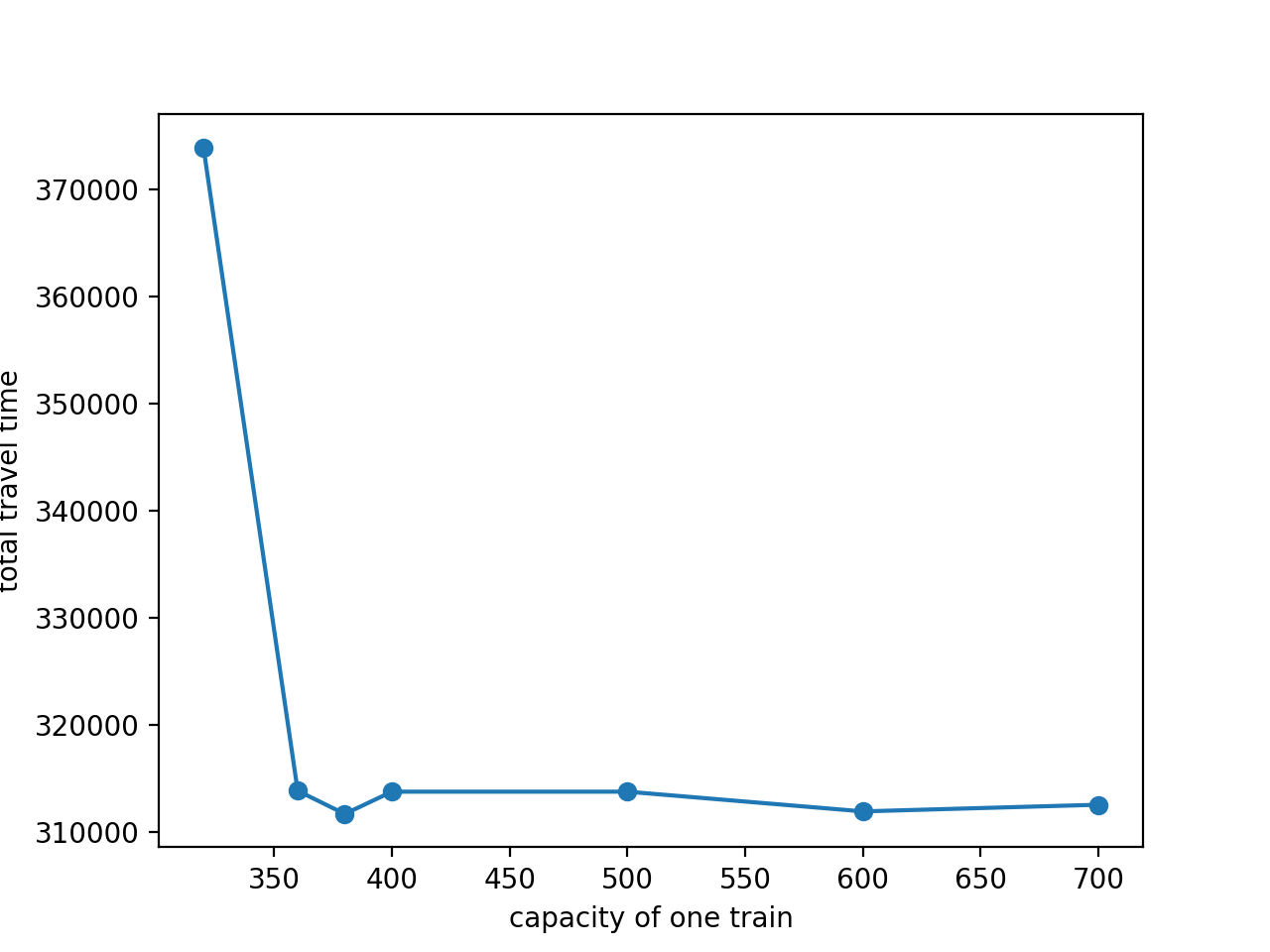} }}
\caption{Sensitivity analysis on capacity}
\label{fig:toy capacity}
\end{figure}

When the capacity of one train is limited, we need to assign more train runs to the system to satisfy all demand. However, with the budget constraint, the number of open lines and train runs are limited, and the objective function (the number of new cases) will increase when we have smaller capacity for each train.

The trend of the new cases and total travel time are similar to what happens when we change budget: there is a minimum capacity where this problem become infeasible, and both of the new cases and total passenger travel are more sensitive when the capacity gets closer to the minimum capacity.

We change the capacity of one train in this model to see how the new case will change when we have different level of capacity (see fig \ref{fig:toy capacity}). For each different parameter, we set the budget level to be 60\% and did 1000 iterations of our LR algorithm and use the best upper bound as the result. In our experiments, we can see that when capacity is greater or equal to 400, the objective function is the same. While when capacity become less than 320, there is still no feasible solution found after 1000 LR iterations, so the transition point is $<320$. The trend of total travel time is similar, but with some fluctuations since the total travel time is not our objective function but is just positively correlated to the objective (new cases), and we are only using approximated algorithms here.

Also note that the transition point for budget and capacity are correlated. If the budget is smaller then the transition point for capacity is greater. For example, we also did the sensitivity analysis when the budget level is 100\%. In this case, the transition point for budget is less than 110, which is much smaller than 320.

\subsection{A real world example: BART}
\label{S: bart}
\subsubsection{Parameters}
\label{S: bart parameter}
BART is a rapid transit system that serves the San Francisco Bay Area. It spans 121 miles of double track, consisting of 48 stations\cite{BART2019}. We consider the peak hours operation (7:00–10:00) as the time horizon.

The network is shown in figure \cite{BART2020a}. The time horizon is discretized into 1-minute intervals. According to the timetable, it can be observed that the in-vehicle travel time is much longer than the dwell time at each station, so the assumptions are satisfied. There are many parallel links in BART, and transfers are prevalent in this system. The proposed space-time network model can capture all these network features.

The operation cost can be estimated from the BART budget report \cite{BART2020a}. The total operating cost budget is 767.8M in 2020. As the operation time of each train dispatch for each line can be calculated, assuming that the operation cost is proportional to the operating time, the cost of one dispatch can be estimated, as shown in Table \ref{tab:line cost}.

Since the demand originated at each station is non-zero, and the walking distance between two stations in BART system is long. We know that we have to open all the lines otherwise the system would be infeasible, since we have strict demand constraints. We set the time tolerance for each user to be $TL=45 min$.

\begin{figure}[h]
    \centering
    \includegraphics[width=0.8\linewidth]{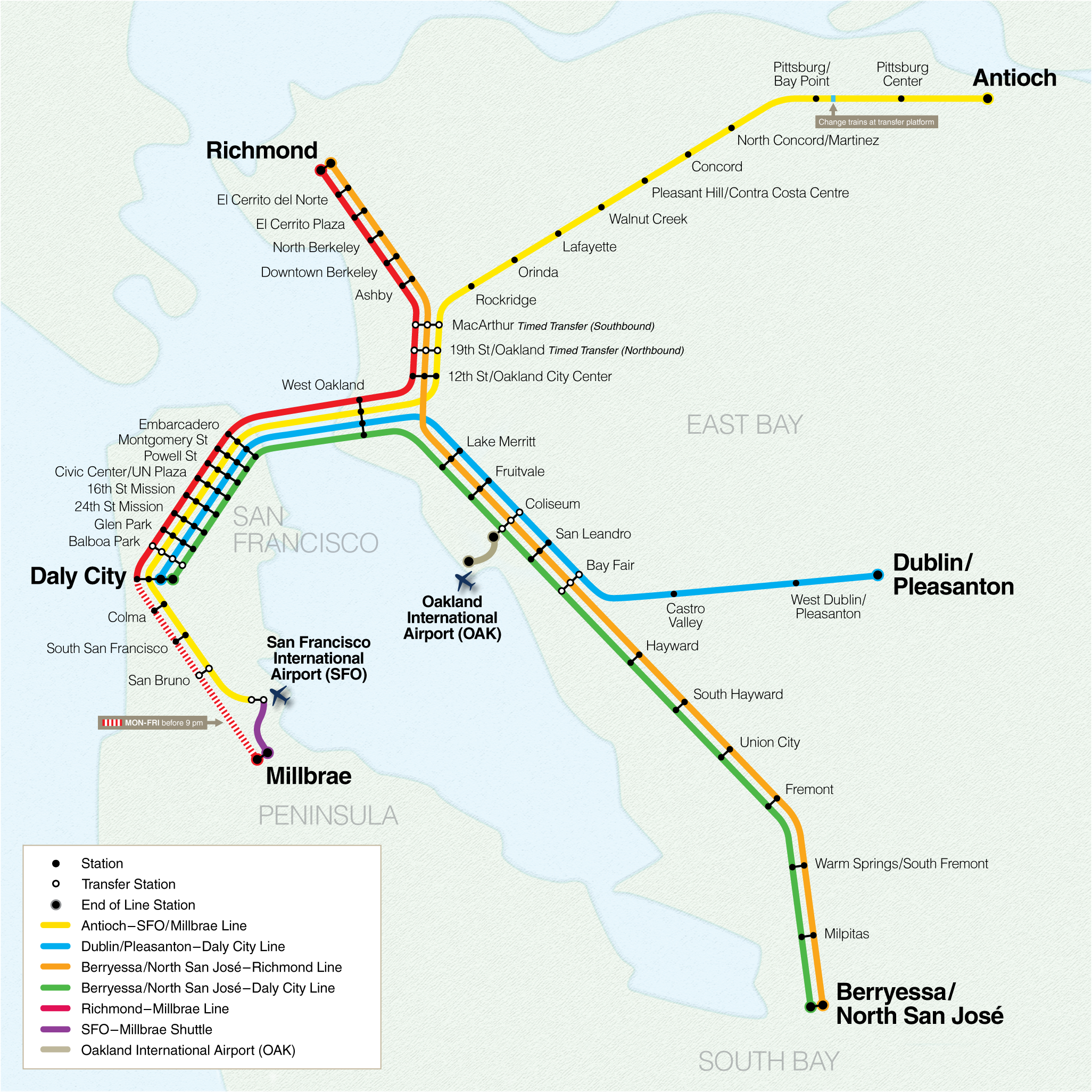}
    \caption{map of BART \cite{BART2020a}}
    \label{fig:BART}
\end{figure}

\begin{table}[h]
\begin{tabular}{c c c c}
\hline
Line number & Color & Line name & Cost(\$)\\
\hline
0&gray & Oakland Airport to Coliseum & 848  \\
1&gray & Coliseum to Oakland Airport & 848  \\               
2&blue & Daly City to Dublin/Pleasanton & 4179 \\
3&blue & Dublin/Pleasanton to Daly City & 4118 \\
4&green & Daly City to Berryessa/North San Jose & 5390 \\              
5&green & Berryessa/North San Jose to Daly City & 5330 \\
6&orange &Berryessa/North San Jose to Richmond & 4966 \\                7&orange &Richmond to Berryessa/North San Jose& 5330 \\
8&purple &Millbrae to SFIA & 545  \\ 
9&purple &SFIA to Millbrae & 545  \\
10&red &Millbrae/Daly City to Richmond & 4361 \\                
11&red & Richmond to Daly City/Millbrae & 4542 \\
12&yellow &Millbrae/SFIA to Antioch& 6299\\               
13&yellow & Antioch to SFIA/Millbrae & 6481\\
\hline
\end{tabular}
\caption{Line info and cost of one run for different lines}
    \label{tab:line cost}
\end{table}

During the pandemic, BART reduced operation in 2020, and now the system is being reopened according to the guidelines \cite{BART2020b}. Disinfectants are being sprayed on the surfaces of train cars and station platforms, and face coverings are required at all times for all riders aged 13 and above. According to BART, the capacity of each car is 30 people under the 6-feet social distancing rule. As long trains are in use during the pandemic, each train's capacity is 600 people. As a platform's capacity is approximately 1.5 times the train capacity, we use 900 as the platform capacity.

In the current timetable, lines 0 and 1 have high operation frequencies as these two lines are Oakland International Airport shuttles that current employ smaller carts called AirBART. Therefore, the operation pattern is different from the other line and we will ignore the these two lines from Oakland Airport and relocate the demand from Oakland Airport to Coliseum station instead.

Hourly OD demand data from the BART API \cite{BART2020c} are used to estimate the intensity of the Poisson flow during the decision time horizon. The demand is aggregated every 20 minutes to reduce the dimensions of the problem. That is, the desired departure time in our time dependent OD matrix must be multiplies of 20 minute, like 0,20,40,60 etc. Here, we consider operation from 8:00AM to 11:00AM. Using the proposed algorithm, the following iteration process is executed. To further simplify the model, train dispatch is only allowed every 10 min, that is, dispatch occurs only at time 8:10, 8:20, and so on. 

\subsubsection{Optimal timetable design and performance of our algorithm}
\label{S:bart iter}
After plugging the BART network in our model and simplifying the problem by only considering a max frequency of 10 minutes and ignoring Oakland Airport line, we got $277$ 0-1 variables. The number of integer variable $x_{lt}$ is $12\times6\times3=216$ since we have 12 lines during a 3-hour time period. And we have one variable for each line and each station.

The number of continuous variables is exponentially many, but we only generate a few of them in our iteration. The network has $22630$ arcs: We have $3888$ links in $\AAA_{travel}$, and $2398$ links in $\AAA_{wait}$ using sparse storage of network by only considering time step where there could be an arrival/departure of a train at the platform. Most of the lines will share platforms except at station MacArthur, 19th St Oakland, Coliseum and Balboa Park, we have $71+53+107+125=356$ links in $\AAA_{transfer}$ where $71,53,107$ and $125$ corresponds to MacArthur, 19th St Oakland, Coliseum and Balboa Park respectively. In addition to these, we have $13000$ links in $\AAA_{os}$ and $2988$ links in $\AAA_{sd}$. Noting that a large number of links in $\AAA_{os}$ and $\AAA_{sd}$ will not make the algorithm much more difficult to solve since they will not add too much route choice, and we use the path-flow formulation and column generation, which only requires doing one shortest path for each OD pair.

The  number of non-zero OD pair is $2800$ as we aggregate demand every $20 min$ and most of the time dependent OD matrix is zero. When solving SUB1, we need to find shortest paths for $2800$ OD pairs. We will keep generate paths until all paths satisfying boundedly rational conditions with negative reduced cost is found, so that the number of basis expands quickly. When most of the capacity constraints are not tight, and the network structure is not too complicated like BART, we can find the optimal solution after less than 100 column generation steps. The typical number of columns needed is less than $15000$ in our iterations and we only have at most $|\AAA_{travel}|+|\WWW|<5000$ constraints (rows) while most of the capacity constraints are not tight.

We have to admit that if the demand level is high and most of the constraints are tight, this problem can be difficult to solve and we cannot find an feasible solution after many LR iteration steps. This is a limit of our problem. But luckily, for BART system, only some trains reaches their capacity like trains running on cross-bay link, where there is a large flow from east bay to the downtown SF every workday morning.

After 1000 LR iterations, the gap is approximately 5.4\%. (LB: 0.470605, UB: 0.496054). The upper bound solution is better than the cost calculated using the current running timetable (cost: 0.500649), which is calculated based on timetable of BART weekday operation\cite{BART2020b} using our column generation algorithm to find the optimal network flow and cost. 

In the proposed optimal solution, all train lines and stations will be opened. The reason is that we need to cover all stations to have a feasible solution and we must open most of lines in BART system to make sure all stations are covered. Although passengers can walk directly from the origins to stations, the walking distance is great since the distance between two station is quite large in BART system. Therefore, passengers using the system must have access to a station nearby, and we need to open most of the lines to cover all stations. Another reason is given by experiments in section \ref{S:bart sa}, the optimal design will open all lines once all stations are covered and the timetable become more flexible to reduce the number of new infections.

The optimal timetable is shown in table \ref{tab:bart timetable}. For example if $x_{2,10}=1$ then we will have one train run dispatched for line 2 at $time=10 min$ ($8:10$). It can be seen that most of the trains will have two to four runs per hour in our optimal plan, while the current timetable has only two runs. It can also be observed that lines 11 and 13 (Richmond to Millbrae, and Antioch to SFIA) has much more runs than lines 10 and 12 ( Millbrae to Richmond, and SFIA to Antioch). During the morning peak hour, people need to leave their homes in the East Bay for travel to the San Francisco CBD area, and our design captured this demand pattern.

Compared to the original timetable, our timetable has higher frequency in the first two hours (8:00AM-10:00AM) or (0-120 min). This is because the demand in the first two hours are much higher than the demand from 10:00AM to 11:00AM. In BART data, the total daily demand is 2200 in 8:00-9:00 AM, 1649 in 9:00-10:00AM and 1337 in 10:00-11:00AM.

Another major difference between our timetable and the existing timetable is that we have much fewer operations for line 8 and 9 (Millbrae to SFIA and SFIA to Millbrae). There are two reasons for this: (1). These lines serves only passengers between Millbrae and SFIA, and the number of passengers is small compared to the whole system ($\approx 0.1\%$, with 14.75 passengers from MLBR to SFIA and 15.57 passengers from SFIA to MLBR while the whole system has an average of 27677.38 daily users). (2). Our algorithm favors long distance lines since we assume all trains need to be cleaned after one operation, and the cleaned cost and time is fixed. Therefore, the per-mile cost is higher for short lines compared to other lines, and the algorithm tend to operate long distance lines.

\begin{figure}[H]
    \centering
    \includegraphics[width=0.8\linewidth]{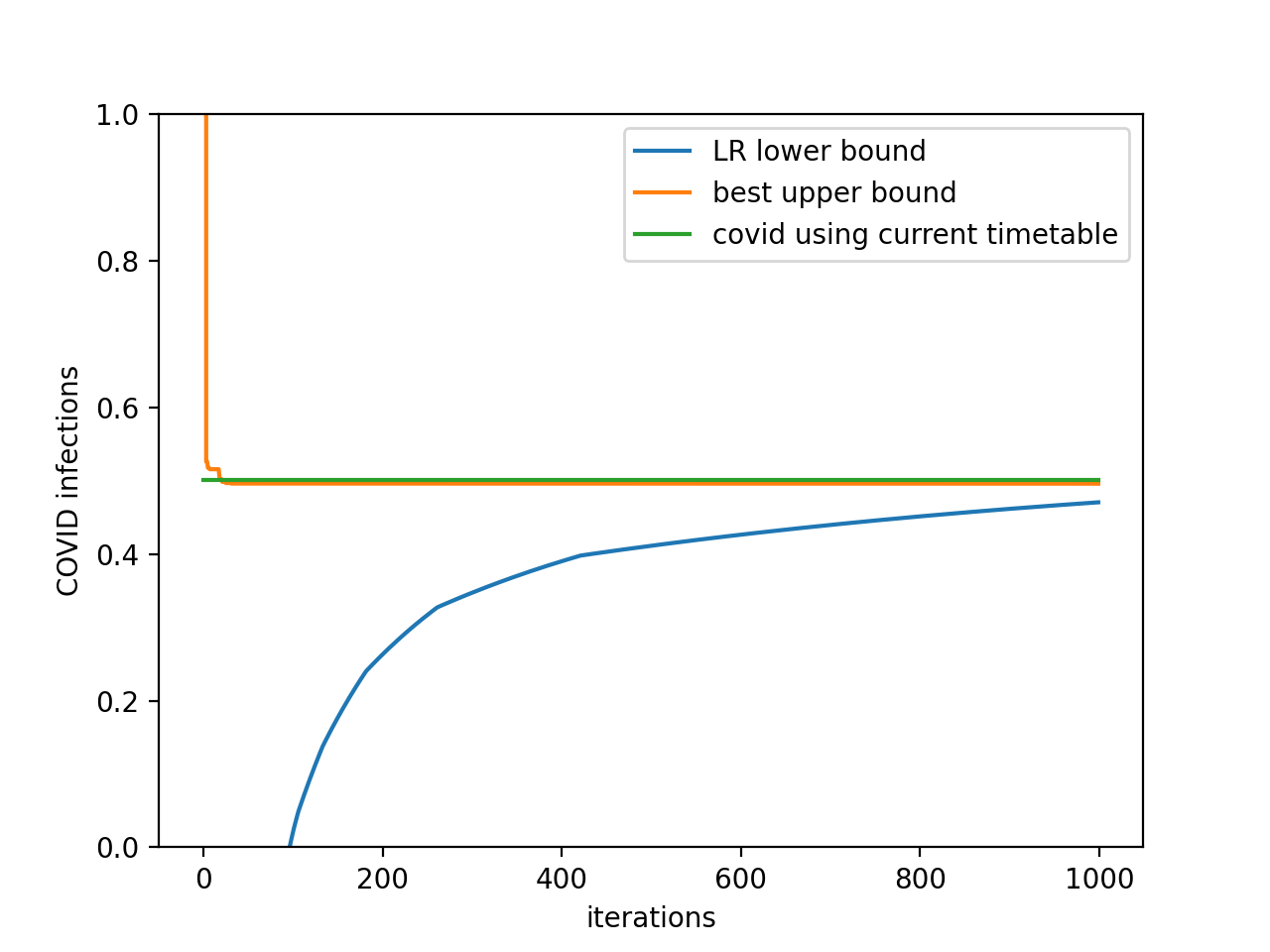}
    \caption{LR iteration of BART example}
    \label{fig:bart iter}
\end{figure}

\begin{table}[H]
    \centering
    \resizebox{\textwidth}{!}{
    \begin{tabular}{c c c c}
    \hline
         Line number & Line name & Optimal timetable & Current timetable \\
    \hline
         2 & Daly City to Dublin/Pleasanton & 0,40,60,70,80,110,120,150 & 21,51,81,111,141,171 \\
         3 & Dublin/Pleasanton to Daly City & 10,30,40,70,80,100,120,140 & 9,39,69,99,129,159\\
         4 & Daly City to Berryessa/North San Jose & 10,30,40,60,90,110,130 & 12,42,72,102,132,162\\
         5 & Berryessa/North San Jose to Daly City &0,20,40,70,100 & 0,30,60,90,120,150\\
         6 & Berryessa/North San Jose to Richmond &20,40,60,70,90,100,170 & 12,42,72,102,132,162\\
         7 & Richmond to Berryessa/North San Jose & 20,30,50,90,110,120,140,150 & 3,33,63,93,123,153\\
         8 & Millbrae to SFIA & 40,110 & 27,57,87,117,147,177\\
         9 & SFIA to Millbrae & 120,170 & 9,39,69,99,109,129,159\\
         10 & Millbrae/Daly City to Richmond & 20,50,70,100 & 18,48,78,108,138,168\\
         11 & Richmond to Daly City/Millbrae & 10,30,40,50,80,90,100,130,140,170 & 9,39,69,99,129,159\\
         12 & Millbrae/SFIA to Antioch &20,50,60,70,80,110,120,170 & 24,54,84,114,144,174\\
         13 & Antioch to SFIA/Millbrae & 10,20,60,70,80,90,100,120,140,150,170 & 0,30,60,90,120,150\\
    \hline
    \end{tabular}
    }
    \caption{Optimal timetable}
    \label{tab:bart timetable}
\end{table}

\subsubsection{What will happen if we close/open lines}
\label{S:bart sa}
From the previous example, we know with the boundedly rational constraints, the problem can be infeasible if we don't cover all the stations, and given the stations are covered, we will open all lines in the optimal solution. To find out why it is the case and what will happen if we don't open all the lines or stations. We assume in this experiment that passengers will go to the closest station from their origin and leave the system from the station closest to their destination using the shortest path in the physical network, so the total demand is the same. We open the lines one by one following table \ref{tab:bart reopen}. The proposed algorithm is then executed for 1000 LR iterations to find the upper/lower bound of the COVID-19 cost, optimal timetable design, and network flow assignment for each experiment. The resulting objective function (new cases), total travel time inside and outside the transit system are shown in table \ref{tab:bart sa}.

\begin{table}[H]
    \centering
    \begin{tabular}{c c c c c c c}
    \hline
         Experiment number & Yellow & Blue & Green & Orange & Red & Purple  \\
    \hline
    0 & \checkmark & & & & &  \\
    1 & \checkmark & \checkmark & & & & \\
    2 & \checkmark & \checkmark & \checkmark & & & \\
    3 & \checkmark & \checkmark & \checkmark & \checkmark & &\\
    4 & \checkmark & \checkmark & \checkmark & \checkmark & \checkmark &\\
    5 & \checkmark & \checkmark & \checkmark & \checkmark & \checkmark & \checkmark\\
    \hline
    \end{tabular}

    \caption{Line opened in each experiment}
    \label{tab:bart reopen}
\end{table}

\begin{table}[H]
\centering
    \resizebox{\textwidth}{!}{
\begin{tabular}{c c c c c c c}
\hline
Experiment number & 0       & 1       & 2       & 3       & 4      & 5       \\
\hline
New cases              & infeasible & 0.42663942 & 0.46613135 & 0.61159565 & 0.5661387 & 0.49605433 \\
Travel time in the system  & infeasible & 60734      & 66413      & 86091      & 80032     & 70073      \\
Travel time outside the system & infeasible & 14134      & 9057       & 336        & 216       & 0          \\
Line 2 runs         & infeasible & 14         & 12         & 13         & 6         & 8          \\
Line 3 runs         & infeasible & 15         & 12         & 13         & 2         & 9          \\
Line 4 runs         & infeasible & 0          & 9          & 2          & 10        & 9          \\
Line 5 runs         & infeasible & 0          & 12         & 3          & 13        & 5          \\
Line 6 runs         & infeasible & 0          & 0          & 13         & 13        & 7          \\
Line 7 runs         & infeasible & 0          & 0          & 3          & 10        & 8          \\
Line 8 runs         & infeasible & 0          & 0          & 0          & 0         & 2          \\
Line 9 runs         & infeasible & 0          & 0          & 0          & 0         & 2          \\
Line 10 runs         & infeasible & 0          & 0          & 0          & 4         & 4          \\
Line 11 runs         & infeasible & 0          & 0          & 0          & 4         & 11         \\
Line 12 runs         & infeasible & 17         & 14         & 14         & 2         & 9          \\
Line 13 runs         & infeasible & 17         & 17         & 7          & 15        & 11 \\  
\hline
\end{tabular}}
\caption{What happens when we close some lines}
\label{tab:bart sa}
\end{table}

From table \ref{tab:bart sa}, we can see when we only open two lines, the demand is huge and we cannot satisfy all the demand in experiment 0. As we expand the number of lines, more stations are covered, the travel cost outside the system (time it takes for passengers to travel from origins to stations and from stations to destinations) will decrease, and the number of train runs needed for each line decreases since we don't rely solely one these lines after opening new lines. Also, we notice that the new cases or travel cost inside the system is not monotone. It depends on the network structure. 
\begin{enumerate}
    \item When we only open a small fraction of lines, not all stations are covered and many people spend most of their time walking outside the system. Therefore, the number of new infection inside the system is small. At this time, opening a new line result in a increase in demand with respect to passenger distance traveled, and the objective function (new cases) will also increase.
    
    \item When we have the most of the network opened, all the stations are covered and we have many parallel lines. These features enable us to make flexible timetables and the optimal new infection number is small. At this time, if we close some existing lines and stations, passengers will go to nearby opened stations and the number of queuing passengers waiting will increase, leading to more infections. In addition, if we close parallel lines in the system, the designed capacity between two stations are less flexible and leads to an increase in both total travel time and the waiting time on platforms, thus increasing the number of infections. This also explained why all lines are opened in the optimal design from section \ref{S:bart iter}
\end{enumerate}

\section{Conclusions and future research}
\label{S:6}
In this study, we combined a simplified SCME model and a space-time dynamic transit network to estimate the impact of the pandemic on the public transit system. The proposed model can predict new the pandemic cases in any transit system. We also developed an algorithm to find the optimal network reopening plan and timetable simultaneously, considering the risk of pandemic infection and social distancing rules. The proposed model and the Lagrangian relaxation frame can also be extended to other economics problems where reopening too quickly under pandemic conditions may result in a 'second wave' of the pandemic.

Based on the numerical example, it can be observed that the proposed algorithm can provide a relatively good lower and upper bounds for the design problem. However, we also notice that this algorithm can only solve relative simple problem where not too many capacity constraints are tight, otherwise the column generation will take a long time and the number of variables explodes.

From the experiments, we know that the sensitivity of budget and capacity constraints are correlated. With capacity fixed, there is a transition point for budget where the problem become infeasible if budget is smaller than this point, and both new infection number and travel time are more sensitive as budget gets closer to this transition point. Similar phenomena holds for capacity with budget fixed. In addition, with more budget, the transition point of capacity become smaller and the number of new cases is less sensitive with respect to capacity.

We also know that if we assume people who need the public transportation system will go to the closest opened station and keep using the system. Then we should either (1) open as many lines as possible and operate parallel lines, so that the system has enough flexibility to deal with time and space dependent travel demand and control the spread of the disease, or (2) only operate a small number of lines so that people are forced outside the system to prevent infections. It is not wise to cover many stations with only small number of lines. In that case, we will keep most of the demand in the system while we cannot deal with this demand with many lines closed.

Our future research will consider the dynamics in stochastic demand modeling. In the current model, the OD demand is deterministic and must be given in constraints \ref{E:flow conservation}. This means that if one OD pair relies on one particular line, then this line must be opened with the optimal design. In contrast, some demand may be sacrificed to decrease the risk of the pandemic. Furthermore, the demand itself is a stochastic process (possibly Markovian) following specific transition models. It is possible to design a non-myopic network reopening plan for each period during the entire reopening phase.

\section*{Acknowledgement}
This research is funded by ITS-Berkeley SB 1 (Project ID: 2021-09)
Conflicts of interest: None



%% References

\clearpage
\bibliographystyle{elsarticle-harv}
%% New version of the num-names style
%\bibliographystyle{elsarticle-num-names}
\bibliography{sample.bib}

\begin{thebibliography}{47}
\expandafter\ifx\csname natexlab\endcsname\relax\def\natexlab#1{#1}\fi
\providecommand{\url}[1]{\texttt{#1}}
\providecommand{\href}[2]{#2}
\providecommand{\path}[1]{#1}
\providecommand{\DOIprefix}{doi:}
\providecommand{\ArXivprefix}{arXiv:}
\providecommand{\URLprefix}{URL: }
\providecommand{\Pubmedprefix}{pmid:}
\providecommand{\doi}[1]{\href{http://dx.doi.org/#1}{\path{#1}}}
\providecommand{\Pubmed}[1]{\href{pmid:#1}{\path{#1}}}
\providecommand{\bibinfo}[2]{#2}
\ifx\xfnm\relax \def\xfnm[#1]{\unskip,\space#1}\fi
%Type = Techreport
\bibitem[{Acemoglu et~al.(2020)Acemoglu, Chernozhukov, Werning and
  Whinston}]{Acemoglu2020}
\bibinfo{author}{Acemoglu, D.}, \bibinfo{author}{Chernozhukov, V.},
  \bibinfo{author}{Werning, I.}, \bibinfo{author}{Whinston, M.D.},
  \bibinfo{year}{2020}.
\newblock \bibinfo{title}{A multi-risk SIR model with optimally targeted
  lockdown}.
\newblock \bibinfo{type}{Report} \bibinfo{number}{0898-2937}. National Bureau
  of Economic Research.
%Type = Misc
\bibitem[{{ACtransit}(2020)}]{AC2020}
\bibinfo{author}{{ACtransit}}, \bibinfo{year}{2020}.
\newblock \URLprefix
  \url{http://www.actransit.org/2020/04/17/ac-transit-service-updates-related-to-the-coronavirus-covid-19/}.
%Type = Book
\bibitem[{Allen et~al.(2008)Allen, Brauer, Van~den Driessche and
  Wu}]{Allen2008}
\bibinfo{author}{Allen, L.J.}, \bibinfo{author}{Brauer, F.},
  \bibinfo{author}{Van~den Driessche, P.}, \bibinfo{author}{Wu, J.},
  \bibinfo{year}{2008}.
\newblock \bibinfo{title}{Mathematical epidemiology}. volume
  \bibinfo{volume}{1945}.
\newblock \bibinfo{publisher}{Springer}.
%Type = Techreport
\bibitem[{Alvarez et~al.(2020)Alvarez, Argente and Lippi}]{Alvarez2020}
\bibinfo{author}{Alvarez, F.E.}, \bibinfo{author}{Argente, D.},
  \bibinfo{author}{Lippi, F.}, \bibinfo{year}{2020}.
\newblock \bibinfo{title}{A simple planning problem for covid-19 lockdown}.
\newblock \bibinfo{type}{Report} \bibinfo{number}{0898-2937}. National Bureau
  of Economic Research.
%Type = Article
\bibitem[{An and Lo(2016)}]{An2016}
\bibinfo{author}{An, K.}, \bibinfo{author}{Lo, H.K.}, \bibinfo{year}{2016}.
\newblock \bibinfo{title}{Two-phase stochastic program for transit network
  design under demand uncertainty}.
\newblock \bibinfo{journal}{Transportation Research Part B: Methodological}
  \bibinfo{volume}{84}, \bibinfo{pages}{157--181}.
\newblock \DOIprefix\doi{10.1016/j.trb.2015.12.009}.
%Type = Misc
\bibitem[{{APTA}(2020)}]{APTA2020}
\bibinfo{author}{{APTA}}, \bibinfo{year}{2020}.
\newblock \URLprefix
  \url{https://www.apta.com/public-transit-response-to-coronavirus/}.
%Type = Misc
\bibitem[{{BART}(2020a)}]{BART2020b}
\bibinfo{author}{{BART}}, \bibinfo{year}{2020}a.
\newblock \URLprefix
  \url{https://www.bart.gov/news/articles/2020/news20200526}.
%Type = Misc
\bibitem[{{BART}(2020b)}]{BART2020c}
\bibinfo{author}{{BART}}, \bibinfo{year}{2020}b.
\newblock \URLprefix \url{https://www.bart.gov/about/reports/ridership}.
%Type = Misc
\bibitem[{{BART}(2021)}]{BART2020a}
\bibinfo{author}{{BART}}, \bibinfo{year}{2021}.
\newblock \URLprefix \url{(http://api.bart.gov/}.
%Type = Techreport
\bibitem[{{Bay Area Rapid Transit (BART)}(2019)}]{BART2019}
\bibinfo{author}{{Bay Area Rapid Transit (BART)}}, \bibinfo{year}{2019}.
\newblock \bibinfo{title}{San Francisco Bay Area Rapid Transit District Adopted
  Budget Fiscal Year 2020}.
\newblock \bibinfo{type}{Report}. BART.
\newblock \URLprefix \url{https://www.bart.gov/sites/default/files/docs}.
%Type = Article
\bibitem[{Birge et~al.(2020)Birge, Candogan and Feng}]{Birge2020}
\bibinfo{author}{Birge, J.R.}, \bibinfo{author}{Candogan, O.},
  \bibinfo{author}{Feng, Y.}, \bibinfo{year}{2020}.
\newblock \bibinfo{title}{Controlling epidemic spread: Reducing economic losses
  with targeted closures} .
%Type = Book
\bibitem[{Brauer et~al.(2019)Brauer, Castillo-Chavez and Feng}]{Brauer2019}
\bibinfo{author}{Brauer, F.}, \bibinfo{author}{Castillo-Chavez, C.},
  \bibinfo{author}{Feng, Z.}, \bibinfo{year}{2019}.
\newblock \bibinfo{title}{Mathematical models in epidemiology}.
\newblock \bibinfo{publisher}{Springer}.
%Type = Article
\bibitem[{Cancela et~al.(2015)Cancela, Mauttone and Urquhart}]{Cancela2015}
\bibinfo{author}{Cancela, H.}, \bibinfo{author}{Mauttone, A.},
  \bibinfo{author}{Urquhart, M.E.}, \bibinfo{year}{2015}.
\newblock \bibinfo{title}{Mathematical programming formulations for transit
  network design}.
\newblock \bibinfo{journal}{Transportation Research Part B: Methodological}
  \bibinfo{volume}{77}, \bibinfo{pages}{17--37}.
\newblock \DOIprefix\doi{10.1016/j.trb.2015.03.006}.
%Type = Misc
\bibitem[{{CDC}(2020a)}]{CDC2020a}
\bibinfo{author}{{CDC}}, \bibinfo{year}{2020}a.
\newblock \bibinfo{title}{Covid-19: Frequently asked questions}.
\newblock \URLprefix \url{https://www.cdc.gov/coronavirus/2019-ncov/faq.html}.
%Type = Misc
\bibitem[{{CDC}(2020b)}]{CDC2020b}
\bibinfo{author}{{CDC}}, \bibinfo{year}{2020}b.
\newblock \bibinfo{title}{Covid-19: How it spreads}.
\newblock \URLprefix \url{https://www.cdc.gov/coronavirus/2019-ncov/faq.html}.
%Type = Misc
\bibitem[{{CDC}(2020c)}]{CDC2020c}
\bibinfo{author}{{CDC}}, \bibinfo{year}{2020}c.
\newblock \bibinfo{title}{Covid-19: using transportatoin: Public transit,
  rideshares and taxis, micro-mobility devices, and personal vehicles}.
\newblock \URLprefix
  \url{https://www.cdc.gov/coronavirus/2019-ncov/daily-life-coping/using-transportation.html}.
%Type = Article
\bibitem[{Chinazzi et~al.(2020)Chinazzi, Davis, Ajelli, Gioannini, Litvinova,
  Merler, y~Piontti, Mu, Rossi and Sun}]{Chinazzi2020}
\bibinfo{author}{Chinazzi, M.}, \bibinfo{author}{Davis, J.T.},
  \bibinfo{author}{Ajelli, M.}, \bibinfo{author}{Gioannini, C.},
  \bibinfo{author}{Litvinova, M.}, \bibinfo{author}{Merler, S.},
  \bibinfo{author}{y~Piontti, A.P.}, \bibinfo{author}{Mu, K.},
  \bibinfo{author}{Rossi, L.}, \bibinfo{author}{Sun, K.}, \bibinfo{year}{2020}.
\newblock \bibinfo{title}{The effect of travel restrictions on the spread of
  the 2019 novel coronavirus (covid-19) outbreak}.
\newblock \bibinfo{journal}{Science} \bibinfo{volume}{368},
  \bibinfo{pages}{395--400}.
%Type = Article
\bibitem[{Fan et~al.(2018)Fan, Mei and Gu}]{Fan2018}
\bibinfo{author}{Fan, W.}, \bibinfo{author}{Mei, Y.}, \bibinfo{author}{Gu, W.},
  \bibinfo{year}{2018}.
\newblock \bibinfo{title}{Optimal design of intersecting bimodal transit
  networks in a grid city}.
\newblock \bibinfo{journal}{Transportation Research Part B: Methodological}
  \bibinfo{volume}{111}, \bibinfo{pages}{203--226}.
\newblock \DOIprefix\doi{10.1016/j.trb.2018.03.007}.
%Type = Article
\bibitem[{Farahani et~al.(2013)Farahani, Miandoabchi, Szeto and
  Rashidi}]{Farahani2013}
\bibinfo{author}{Farahani, R.Z.}, \bibinfo{author}{Miandoabchi, E.},
  \bibinfo{author}{Szeto, W.Y.}, \bibinfo{author}{Rashidi, H.},
  \bibinfo{year}{2013}.
\newblock \bibinfo{title}{A review of urban transportation network design
  problems}.
\newblock \bibinfo{journal}{European Journal of Operational Research}
  \bibinfo{volume}{229}, \bibinfo{pages}{281--302}.
\newblock \DOIprefix\doi{10.1016/j.ejor.2013.01.001}.
%Type = Misc
\bibitem[{Frost(2020)}]{Frost2020}
\bibinfo{author}{Frost, M.}, \bibinfo{year}{2020}.
\newblock \bibinfo{title}{New york city subway ridership down 92 percent due to
  coronavirus}.
\newblock \URLprefix
  \url{https://brooklyneagle.com/articles/2020/04/08/new-york-city-subway-ridership-down-92-percent-due-to-coronavirus/}.
%Type = Article
\bibitem[{Gao et~al.(2004)Gao, Sun and Shan}]{Gao2004}
\bibinfo{author}{Gao, Z.}, \bibinfo{author}{Sun, H.}, \bibinfo{author}{Shan,
  L.L.}, \bibinfo{year}{2004}.
\newblock \bibinfo{title}{A continuous equilibrium network design model and
  algorithm for transit systems}.
\newblock \bibinfo{journal}{Transportation Research Part B: Methodological}
  \bibinfo{volume}{38}, \bibinfo{pages}{235--250}.
\newblock \DOIprefix\doi{10.1016/s0191-2615(03)00011-0}.
%Type = Article
\bibitem[{Gershon et~al.(2020)Gershon, Lipton and Levine}]{Gershon2020}
\bibinfo{author}{Gershon, D.}, \bibinfo{author}{Lipton, A.},
  \bibinfo{author}{Levine, H.}, \bibinfo{year}{2020}.
\newblock \bibinfo{title}{Managing covid-19 pandemic without destructing the
  economy}.
\newblock \bibinfo{journal}{arXiv} \bibinfo{volume}{10324}.
%Type = Techreport
\bibitem[{Glover et~al.(2020)Glover, Heathcote, Krueger and
  Ríos-Rull}]{Glover2020}
\bibinfo{author}{Glover, A.}, \bibinfo{author}{Heathcote, J.},
  \bibinfo{author}{Krueger, D.}, \bibinfo{author}{Ríos-Rull, J.V.},
  \bibinfo{year}{2020}.
\newblock \bibinfo{title}{Health versus wealth: On the distributional effects
  of controlling a pandemic}.
\newblock \bibinfo{type}{Report} \bibinfo{number}{0898-2937}. National Bureau
  of Economic Research.
%Type = Article
\bibitem[{Guihaire and Hao(2008)}]{Guihaire2008}
\bibinfo{author}{Guihaire, V.}, \bibinfo{author}{Hao, J.K.},
  \bibinfo{year}{2008}.
\newblock \bibinfo{title}{Transit network design and scheduling: A global
  review}.
\newblock \bibinfo{journal}{Transportation Research Part A: Policy and
  Practice} \bibinfo{volume}{42}, \bibinfo{pages}{1251--1273}.
\newblock \DOIprefix\doi{10.1016/j.tra.2008.03.011}.
%Type = Misc
\bibitem[{{GUROBI}(2020)}]{GUROBI2020}
\bibinfo{author}{{GUROBI}}, \bibinfo{year}{2020}.
\newblock \URLprefix \url{http://www.gurobi.com}.
%Type = Article
\bibitem[{Ibarra-Rojas et~al.(2015)Ibarra-Rojas, Delgado, Giesen and
  Muñoz}]{Ibarra2015}
\bibinfo{author}{Ibarra-Rojas, O.J.}, \bibinfo{author}{Delgado, F.},
  \bibinfo{author}{Giesen, R.}, \bibinfo{author}{Muñoz, J.C.},
  \bibinfo{year}{2015}.
\newblock \bibinfo{title}{Planning, operation, and control of bus transport
  systems: A literature review}.
\newblock \bibinfo{journal}{Transportation Research Part B: Methodological}
  \bibinfo{volume}{77}, \bibinfo{pages}{38--75}.
\newblock \DOIprefix\doi{10.1016/j.trb.2015.03.002}.
%Type = Article
\bibitem[{Ives and Bozzuto(2020)}]{Ives2020}
\bibinfo{author}{Ives, A.R.}, \bibinfo{author}{Bozzuto, C.},
  \bibinfo{year}{2020}.
\newblock \bibinfo{title}{State-by-state estimates of r0 at the start of
  covid-19 outbreaks in the usa}.
\newblock \bibinfo{journal}{medRxiv} ,
  \bibinfo{pages}{2020.05.17.20104653}\DOIprefix\doi{10.1101/2020.05.17.20104653}.
%Type = Misc
\bibitem[{{JHU}(2020)}]{JHU}
\bibinfo{author}{{JHU}}, \bibinfo{year}{2020}.
\newblock \URLprefix \url{https://coronavirus.jhu.edu/}.
%Type = Article
\bibitem[{Jia et~al.(2020)Jia, Lu, Yuan, Xu, Jia and Christakis}]{Jia2020}
\bibinfo{author}{Jia, J.S.}, \bibinfo{author}{Lu, X.}, \bibinfo{author}{Yuan,
  Y.}, \bibinfo{author}{Xu, G.}, \bibinfo{author}{Jia, J.},
  \bibinfo{author}{Christakis, N.A.}, \bibinfo{year}{2020}.
\newblock \bibinfo{title}{Population flow drives spatio-temporal distribution
  of covid-19 in china}.
\newblock \bibinfo{journal}{Nature} \bibinfo{volume}{582},
  \bibinfo{pages}{389--394}.
\newblock \URLprefix \url{https://doi.org/10.1038/s41586-020-2284-y},
  \DOIprefix\doi{10.1038/s41586-020-2284-y}.
%Type = Article
\bibitem[{Joselow(2020)}]{Joselow2020}
\bibinfo{author}{Joselow, M.}, \bibinfo{year}{2020}.
\newblock \bibinfo{title}{There is little evidence that mass transit poses a
  risk of coronavirus outbreaks}.
\newblock \bibinfo{journal}{Scientific American} .
%Type = Article
\bibitem[{Kaplan(2020)}]{Kaplan2020}
\bibinfo{author}{Kaplan, E.H.}, \bibinfo{year}{2020}.
\newblock \bibinfo{title}{Covid-19 scratch models to support local decisions}.
\newblock \bibinfo{journal}{Manufacture and Service Operations Management}
  \bibinfo{volume}{22}, \bibinfo{pages}{645--655}.
%Type = Article
\bibitem[{Kumar et~al.(2021)Kumar, Khani, Lind and Levin}]{Kumar2021}
\bibinfo{author}{Kumar, P.}, \bibinfo{author}{Khani, A.},
  \bibinfo{author}{Lind, E.}, \bibinfo{author}{Levin, J.},
  \bibinfo{year}{2021}.
\newblock \bibinfo{title}{Modeling epidemic spreading through public transit
  using time-varying encounter network}.
\newblock \bibinfo{journal}{Transportation Research Record} .
%Type = Article
\bibitem[{Li et~al.(2020)Li, Pei, Chen, Song, Zhang, Yang and Shaman}]{Li2020}
\bibinfo{author}{Li, R.}, \bibinfo{author}{Pei, S.}, \bibinfo{author}{Chen,
  B.}, \bibinfo{author}{Song, Y.}, \bibinfo{author}{Zhang, T.},
  \bibinfo{author}{Yang, W.}, \bibinfo{author}{Shaman, J.},
  \bibinfo{year}{2020}.
\newblock \bibinfo{title}{Substantial undocumented infection facilitates the
  rapid dissemination of novel coronavirus (sars-cov-2)}.
\newblock \bibinfo{journal}{Science} \bibinfo{volume}{368},
  \bibinfo{pages}{489--493}.
%Type = Article
\bibitem[{Liu and Zhou(2016)}]{Liu2016}
\bibinfo{author}{Liu, J.}, \bibinfo{author}{Zhou, X.}, \bibinfo{year}{2016}.
\newblock \bibinfo{title}{Capacitated transit service network design with
  boundedly rational agents}.
\newblock \bibinfo{journal}{Transportation Research Part B: Methodological}
  \bibinfo{volume}{93}, \bibinfo{pages}{225--250}.
\newblock \DOIprefix\doi{10.1016/j.trb.2016.07.015}.
%Type = Article
\bibitem[{Liu et~al.(2020)Liu, Yin, Ma, Zhang and Zhao}]{Liu2020}
\bibinfo{author}{Liu, K.}, \bibinfo{author}{Yin, L.}, \bibinfo{author}{Ma, Z.},
  \bibinfo{author}{Zhang, F.}, \bibinfo{author}{Zhao, J.},
  \bibinfo{year}{2020}.
\newblock \bibinfo{title}{Investigating physical encounters of individuals in
  urban metro systems with large-scale smart card data}.
\newblock \bibinfo{journal}{Physica A: Statistical Mechanics and its
  Applications} \bibinfo{volume}{545}.
%Type = Article
\bibitem[{Luo et~al.(2020)Luo, Lei, Hai, Xiao, Rui, Yang, Jing, Wang, Xie, Luo,
  Li, Li, Tan, Xu, Yang, Hu and Chen}]{Luo2020Hunan}
\bibinfo{author}{Luo, K.}, \bibinfo{author}{Lei, Z.}, \bibinfo{author}{Hai,
  Z.}, \bibinfo{author}{Xiao, S.}, \bibinfo{author}{Rui, J.},
  \bibinfo{author}{Yang, H.}, \bibinfo{author}{Jing, X.},
  \bibinfo{author}{Wang, H.}, \bibinfo{author}{Xie, Z.}, \bibinfo{author}{Luo,
  P.}, \bibinfo{author}{Li, W.}, \bibinfo{author}{Li, Q.},
  \bibinfo{author}{Tan, H.}, \bibinfo{author}{Xu, Z.}, \bibinfo{author}{Yang,
  Y.}, \bibinfo{author}{Hu, S.}, \bibinfo{author}{Chen, T.},
  \bibinfo{year}{2020}.
\newblock \bibinfo{title}{Transmission of sars-cov-2 in public transportation
  vehicles: A case study in hunan province, china}.
\newblock \bibinfo{journal}{Open Forum Infectious Diseases}
  \bibinfo{volume}{7}.
\newblock \URLprefix \url{https://doi.org/10.1093/ofid/ofaa430},
  \DOIprefix\doi{10.1093/ofid/ofaa430},
  \href{http://arxiv.org/abs/https://academic.oup.com/ofid/article-pdf/7/10/ofaa430/34006085/ofaa430.pdf}{{\tt
  arXiv:https://academic.oup.com/ofid/article-pdf/7/10/ofaa430/34006085/ofaa430.pdf}}.
  \bibinfo{note}{ofaa430}.
%Type = Book
\bibitem[{Martcheva(2015)}]{Martcheva2015}
\bibinfo{author}{Martcheva, M.}, \bibinfo{year}{2015}.
\newblock \bibinfo{title}{An introduction to mathematical epidemiology}.
  volume~\bibinfo{volume}{61}.
\newblock \bibinfo{publisher}{Springer}.
%Type = Article
\bibitem[{Mo et~al.(2021)Mo, Feng, Shen, Tam, Li, Yafeng and Jinhua}]{Mo2021}
\bibinfo{author}{Mo, B.}, \bibinfo{author}{Feng, K.}, \bibinfo{author}{Shen,
  Y.}, \bibinfo{author}{Tam, C.}, \bibinfo{author}{Li, D.},
  \bibinfo{author}{Yafeng, Y.}, \bibinfo{author}{Jinhua, Z.},
  \bibinfo{year}{2021}.
\newblock \bibinfo{title}{Modeling epidemic spreading through public transit
  using time-varying encounter network}.
\newblock \bibinfo{journal}{Transportation Research Part C: Emerging
  Technologies} \bibinfo{volume}{122}.
%Type = Article
\bibitem[{Niu and Zhou(2013)}]{Niu2013}
\bibinfo{author}{Niu, H.}, \bibinfo{author}{Zhou, X.}, \bibinfo{year}{2013}.
\newblock \bibinfo{title}{Optimizing urban rail timetable under time-dependent
  demand and oversaturated conditions}.
\newblock \bibinfo{journal}{Transportation Research Part C: Emerging
  Technologies} \bibinfo{volume}{36}, \bibinfo{pages}{212--230}.
\newblock \DOIprefix\doi{10.1016/j.trc.2013.08.016}.
%Type = Article
\bibitem[{Niu et~al.(2015)Niu, Zhou and Gao}]{Niu2015}
\bibinfo{author}{Niu, H.}, \bibinfo{author}{Zhou, X.}, \bibinfo{author}{Gao,
  R.}, \bibinfo{year}{2015}.
\newblock \bibinfo{title}{Train scheduling for minimizing passenger waiting
  time with time-dependent demand and skip-stop patterns: Nonlinear integer
  programming models with linear constraints}.
\newblock \bibinfo{journal}{Transportation Research Part B: Methodological}
  \bibinfo{volume}{76}, \bibinfo{pages}{117--135}.
\newblock \DOIprefix\doi{10.1016/j.trb.2015.03.004}.
%Type = Article
\bibitem[{Sadik-Khan and Solomonow(2020)}]{Sadik2020}
\bibinfo{author}{Sadik-Khan, J.}, \bibinfo{author}{Solomonow, S.},
  \bibinfo{year}{2020}.
\newblock \bibinfo{title}{Fear of public transit got ahead of the evidence}.
\newblock \bibinfo{journal}{The Atlantic} .
%Type = Article
\bibitem[{Sanche et~al.(2020)Sanche, Lin, Xu, Romero-Severson, Hengartner and
  Ke}]{Sanche2020}
\bibinfo{author}{Sanche, S.}, \bibinfo{author}{Lin, Y.T.}, \bibinfo{author}{Xu,
  C.}, \bibinfo{author}{Romero-Severson, E.}, \bibinfo{author}{Hengartner, N.},
  \bibinfo{author}{Ke, R.}, \bibinfo{year}{2020}.
\newblock \bibinfo{title}{High contagiousness and rapid spread of severe acute
  respiratory syndrome coronavirus 2}.
\newblock \bibinfo{journal}{Emerging Infectious Diseases} \bibinfo{volume}{26},
  \bibinfo{pages}{1470--1477}.
\newblock \DOIprefix\doi{https://dx.doi.org/10.3201/eid2607.200282}.
%Type = Misc
\bibitem[{{Sante Publique France}(2020)}]{SPF2020}
\bibinfo{author}{{Sante Publique France}}, \bibinfo{year}{2020}.
\newblock \bibinfo{title}{Covid-19 : point épidémiologique du 4 juin 2020}.
\newblock \URLprefix
  \url{https://www.santepubliquefrance.fr/maladies-et-traumatismes/maladies-et-infections-respiratoires/infection-a-coronavirus/documents/bulletin-national/covid-19-point-epidemiologique-du-4-juin-2020}.
%Type = Article
\bibitem[{Shen et~al.(2020)Shen, Li, Dong, Wang, Martinez, Sun, Handel, Chen,
  Chen, Ebell, Wang, Yi, Wang, Wang, Wang, Chen, Qi, Liang, Li, Ling, Chen and
  Xu}]{Shen2020Zhejiang}
\bibinfo{author}{Shen, Y.}, \bibinfo{author}{Li, C.}, \bibinfo{author}{Dong,
  H.}, \bibinfo{author}{Wang, Z.}, \bibinfo{author}{Martinez, L.},
  \bibinfo{author}{Sun, Z.}, \bibinfo{author}{Handel, A.},
  \bibinfo{author}{Chen, Z.}, \bibinfo{author}{Chen, E.},
  \bibinfo{author}{Ebell, M.H.}, \bibinfo{author}{Wang, F.},
  \bibinfo{author}{Yi, B.}, \bibinfo{author}{Wang, H.}, \bibinfo{author}{Wang,
  X.}, \bibinfo{author}{Wang, A.}, \bibinfo{author}{Chen, B.},
  \bibinfo{author}{Qi, Y.}, \bibinfo{author}{Liang, L.}, \bibinfo{author}{Li,
  Y.}, \bibinfo{author}{Ling, F.}, \bibinfo{author}{Chen, J.},
  \bibinfo{author}{Xu, G.}, \bibinfo{year}{2020}.
\newblock \bibinfo{title}{{Community Outbreak Investigation of SARS-CoV-2
  Transmission Among Bus Riders in Eastern China}}.
\newblock \bibinfo{journal}{JAMA Internal Medicine} \bibinfo{volume}{180},
  \bibinfo{pages}{1665--1671}.
\newblock \URLprefix \url{https://doi.org/10.1001/jamainternmed.2020.5225},
  \DOIprefix\doi{10.1001/jamainternmed.2020.5225},
  \href{http://arxiv.org/abs/https://jamanetwork.com/journals/jamainternalmedicine/articlepdf/2770172/jamainternal\_shen\_2020\_oi\_200076\_1611345809.71283.pdf}{{\tt
  arXiv:https://jamanetwork.com/journals/jamainternalmedicine/articlepdf/2770172/jamainternal\_shen\_2020\_oi\_200076\_1611345809.71283.pdf}}.
%Type = Article
\bibitem[{Szeto and Jiang(2014)}]{Szeto2014}
\bibinfo{author}{Szeto, W.Y.}, \bibinfo{author}{Jiang, Y.},
  \bibinfo{year}{2014}.
\newblock \bibinfo{title}{Transit route and frequency design: Bi-level modeling
  and hybrid artificial bee colony algorithm approach}.
\newblock \bibinfo{journal}{Transportation Research Part B: Methodological}
  \bibinfo{volume}{67}, \bibinfo{pages}{235--263}.
\newblock \DOIprefix\doi{10.1016/j.trb.2014.05.008}.
%Type = Misc
\bibitem[{{The People's Government of Hubei Province}(2020)}]{Hubei2020}
\bibinfo{author}{{The People's Government of Hubei Province}},
  \bibinfo{year}{2020}.
\newblock \bibinfo{title}{Wuhan public transport will officially provide
  commuter services from march 16}.
\newblock \URLprefix
  \url{https://www.hubei.gov.cn/zhuanti/2020/gzxxgzbd/sz/202003/t20200315_2182086.shtml}.
%Type = Article
\bibitem[{Verbas and Mahmassani(2015)}]{Verbas2015}
\bibinfo{author}{Verbas, I.O.}, \bibinfo{author}{Mahmassani, H.S.},
  \bibinfo{year}{2015}.
\newblock \bibinfo{title}{Integrated frequency allocation and user assignment
  in multimodal transit networks}.
\newblock \bibinfo{journal}{Transportation Research Board}
  \bibinfo{volume}{2498}, \bibinfo{pages}{37--45}.
\newblock \URLprefix \url{https://dx.doi.org/10.3141/2498-05},
  \DOIprefix\doi{10.3141/2498-05}.

\end{thebibliography}






% \end{thebibliography}

\end{document}